\documentclass[11pt,a4paper]{article}
\pdfoutput=1
\usepackage{jheppub}
\usepackage{amsfonts,amssymb,amsmath}
\usepackage{epsfig, graphicx,yfonts}

\newcommand{\be}{\begin{equation}}
\newcommand{\ee}{\end{equation}}
\newcommand{\bea}{\begin{eqnarray}}
\newcommand{\eea}{\end{eqnarray}}
\newcommand{\eqn}[1]{(\ref{#1})}

\newcommand{\sac}{\, , \qquad}
\newcommand{\mt}[1]{\textrm{\tiny #1}}
\newcommand{\jem}{J^\mt{EM}}
\newcommand{\cf}{{\cal F}}
\newcommand{\cb}{{\cal B}}
\newcommand{\ch}{{\cal H}}

\newcommand{\vev}[1]{\langle #1\rangle}
\newcommand{\wn}{{\textswab{w}}}
\def\nc {N_\mt{c}}
\def\nf {N_\mt{f}}
\def\uh {u_\mt{H}}
\def\gym {g_\mt{YM}}

\title{Thermal photon production \\  in a  strongly coupled anisotropic plasma}
\author[a]{Leonardo Pati\~no}
\author[b,c]{and Diego Trancanelli} 
\affiliation[a]{Departamento de F\'isica, Facultad de Ciencias, Universidad Nacional Aut\'onoma de M\'exico, A.P. 50-542, M\'exico D.F. 04510, M\'exico}
\affiliation[b]{Instituto de F\'isica, Universidade de S\~ao Paulo, 05314-970 S\~ao Paulo, Brazil} 
\affiliation[c]{Department of Physics, University of Wisconsin, Madison, WI 53706, USA} 
  
\date{\today}

\abstract{
Photons produced in heavy ion collisions escape from the surrounding medium virtually unperturbed by it, thus representing an excellent probe of the conditions at the emission point. Using the gauge/gravity duality, we calculate the rate of photon production in an anisotropic, strongly coupled ${\cal N}=4$ plasma with $\nf\ll \nc$ quark flavors. We consider arbitrary orientations of the photon momentum with respect to the anisotropic direction, as well as arbitrary values of the anisotropy. We present results for the correlation functions of two electromagnetic currents and for the electric conductivity. These quantities can be larger or smaller than the isotropic ones, depending on the direction of propagation and polarization of the photons. The total production rate is however always larger than the isotropic one, independently of the frequency, direction of propagation, and value of the anisotropy.}
  

\emailAdd{leopj@ciencias.unam.mx} 
\emailAdd{dtrancan@fma.if.usp.br} 
  
\begin{document}  

\begin{flushright}
MAD-TH-12-07
\end{flushright}

\maketitle
\setlength{\parskip}{8pt}


\section{Introduction}

Just as the photons in the cosmic microwave background provide us with a detailed snapshot of the conditions of the Universe soon after the Big Bang, photons produced in heavy ion collisions provide excellent probes of the ``Little Bang'' tested at RHIC \cite{rhic,rhic2} and LHC \cite{lhc}. The reason is that, like the Universe after recombination, the quark-gluon plasma (QGP) created in those experiments is optically transparent, both because of its limited spatial extent and because of the smallness of the electromagnetic coupling $\alpha_\mt{EM}$. Photons produced inside this medium escape with essentially no rescattering, bringing with them valuable information about the conditions of the system at the location of their emission \cite{photon,photon1}.

While it is possible to compute the production rate of such photons in a plasma at small $\alpha_\mt{s}$ using ordinary perturbative techniques (see e.g. \cite{Arnold:2001ms}), the QGP produced at RHIC and LHC seemingly behaves as a strongly coupled fluid \cite{fluid,fluid2}, thus rendering a perturbative approach problematic. This non-perturbative nature of the plasma provides a motivation to apply the gauge/gravity correspondence \cite{duality,duality2,duality3}\footnote{See \cite{review} for a review of  applications to the QGP.} to this system. Unfortunately, the holographic dual to QCD is not currently known, but the hope is that studying `reasonable' proxies, such as finite temperature ${\cal N}=4$ super Yang-Mills (SYM),\footnote{The zero-temperature limits of QCD and ${\cal N}=4$ SYM are completely different theories (confining the former, conformal and supersymmetric the latter, among other differences), but at the typical temperatures explored at RHIC and LHC many of these differences disappear; see e.g. \cite{soup}.} 
one might gain insight about the qualitative features of the strong coupling dynamics of the plasma. Moreover, some quantities turn out to be quite model-independent and universal, a notable example being the shear viscosity to entropy density ratio \cite{Kovtun:2004de}.    

The study of photon emission in a strongly coupled ${\cal N}=4$ plasma using holography was initiated in \cite{CaronHuot:2006te}, where photons coupled to massless quarks in the adjoint representation were studied, and continued in \cite{Parnachev:2006ev} and \cite{Mateos:2007yp},  where massless and massive quarks in the fundamental representation were the electrically charged particles. Photon production in a `soft wall' AdS/QCD model was addressed in \cite{Atmaja:2008mt,Bu:2012zza} and in a charged black hole background in \cite{Jo:2010sg}. String corrections to the supergravity results for the electric conductivity and spectral density were studied in \cite{corr1,corr2,corr3}.

The geometry used in those works was spatially isotropic in the gauge theory directions and therefore dual to a rotationally invariant plasma. It is now believed that an important feature of the real-world QGP is the fact that the pressures along the beam direction and along the transverse directions are unequal in the initial stages of the plasma evolution and that the system displays a sizable initial anisotropy \cite{anis1,anis2}. A first analysis of the electromagnetic signatures in a strongly coupled anisotropic plasma was carried on in \cite{Rebhan:2011ke}, where the geometry of \cite{Janik:2008tc} was used as dual. An undesirable feature of that model was the presence of a naked singularity, that nonetheless allowed the imposition of infalling boundary conditions and the definition of retarded correlators. 

In this paper we revisit the computation of photon production in a strongly coupled anisotropic plasma using as gravitational dual the type IIB supergravity solution that was discovered in \cite{Mateos:2011ix,Mateos:2011tv}, building on inspiration from \cite{ALT}.  This solution is static and corresponds to a plasma in anisotropic thermal equilibrium, with the force responsible for maintaining the equilibrium being a position-dependent theta-angle, as we shall review below. Of course, a real-world plasma undergoes a non-trivial time evolution, but the assumption of staticity might be motivated by considering time scales much shorter than the characteristic evolution scale, which is as long as a few fm/c (see e.g. \cite{fmoverc}). 

The geometry of \cite{Mateos:2011ix,Mateos:2011tv} enjoys several nice features that make it an ideal toy model for addressing, from first principles, the effects of spatial anisotropy on physical observables of interest. Such features include regularity of the fields on and outside the horizon, asymptotic AdS boundary conditions, and a UV completion with a solid embedding in type IIB string theory. We introduce $\nf\ll \nc$ flavors of massless quarks in this background and compute the production rate of photons for arbitrary orientations of the photon wave vectors and for arbitrary values of the anisotropy. We also compute the electric DC (i.e. zero-frequency) conductivity and compare it with that of an isotropic plasma which is kept either at the same temperature or at the same entropy density. We find that the conductivity can be larger or smaller than the isotropic conductivity, depending on the orientation of the photon polarization with respect to the anisotropic direction. The total production rate, on the other hand, is always larger than the isotropic rate, independently of the frequency of the photons, their direction of propagation, and the value of the anisotropy. Moreover, the larger the degree of anisotropy of the plasma, the more brightly this glows compared to the isotropic case.

Other physical properties of the anisotropic geometry considered here have already been studied in the literature. These include the shear viscosity to entropy density ratio \cite{rebhan_viscosity,mamo}, the drag force experienced by a heavy quark \cite{Chernicoff:2012iq,giataganas}, the energy lost by a quark rotating in the transverse plane \cite{fadafan}, the stopping distance of a light probe \cite{stopping}, the jet quenching parameter of the medium \cite{giataganas,Rebhan:2012bw,jet}, and the potential between a quark and antiquark pair, both static \cite{giataganas,Rebhan:2012bw,Chernicoff:2012bu} and in a plasma wind \cite{Chernicoff:2012bu}. A non-interacting version of the model of \cite{Mateos:2011ix,Mateos:2011tv} has also been considered in \cite{rebhan3} and the corresponding phase diagram has been obtained.

The structure of the paper is as follows. In Sec.~\ref{sec1} we derive the field theoretical expression for the photon production rate in the presence of anisotropy. In Sec.~\ref{sec2} we review the gravitational dual to the anisotropic system we consider and describe the introduction of flavor branes in this background. In Sec.~\ref{sec3} we compute the photon production rate and conductivity from holography and present our results. We conclude with a discussion and an outlook in Sec.~\ref{sec4}.


\section{Photon production in an anisotropic plasma}
\label{sec1}

The gauge theory we shall consider is obtained via an isotropy breaking deformation of four-dimensional ${\cal N}=4$ super Yang-Mills (SYM) with gauge group $SU(\nc)$, at large $\nc$ and large 't~Hooft coupling $\lambda=\gym^2\nc$. The deformation consists in including in the action a theta-term which depends linearly on one of the spatial directions, say $z$, \cite{ALT}
\be
S_{SU(\nc)}=S_{{\cal N}=4}+\frac{1}{8\pi^2}\int \theta(z)\, \mathrm{Tr}F\wedge F\,,
\qquad 
\theta(z)\propto z \,,
\label{andef}
\ee
where the proportionality constant in $\theta(z)$ has dimensions of energy and will be related to the parameter $a$ that we shall introduce in the next section. The rotational $SO(3)$ symmetry in the space directions is broken by the new term down to $SO(2)$ in the $xy$-plane. For this reason we shall call the $z$-direction the longitudinal (or anisotropic) direction, while $x$ and $y$ will be the transverse directions. 

This theory has matter fields in the adjoint representation of the gauge group. We can also introduce $\nf$ flavors of scalars $\Phi^a$ and fermions $\Psi^a$ in the fundamental representation, with the index $a$ taking values between 1 and $\nf$. With an abuse of language, we will refer to these fundamental fields indistinctly as `quarks'. 

To study photon production we turn on a dynamical photon by including a $U(1)$ kinetic term in the action (\ref{andef}) and a coupling to the fields that we want to be charged under this Abelian symmetry. In order to realize a situation as similar to QCD as possible, we require that only the fundamental fields be charged, while the adjoint fields are to remain neutral. The action for the resulting $SU(\nc)\times U(1)$ theory is then
\be
S = S_{SU(\nc)} -\frac{1}{4}\int d^4x\left(  F_{\mu \nu}^2 - 4e \, A^\mu J^\mt{EM}_\mu\right) \,,
\label{fulllagr}
\ee
where $F_{\mu\nu} = \partial_\mu A_\nu - \partial_\nu A_\mu$ is the $U(1)$ gauge field strength, $e$ the electromagnetic coupling, and the electromagnetic current is given by 
\be
J^\mt{EM}_\mu = \bar{\Psi} \gamma_\mu \Psi 
+ \frac{i}{2} \Phi^* \left( {\cal D}_\mu \Phi \right) 
- \frac{i}{2} \left( {\cal D}_\mu \Phi \right)^* \Phi \,,
\label{current}
\ee
with an implicit sum over the flavor indices. The fundamental fields are differentiated using the full $SU(\nc)\times U(1)$ connection, ${\cal D}_\mu=D_\mu-ieA_\mu$, while the $SU(\nc)$-covariant derivative $D_\mu$ acts on the adjoint matter. 

\begin{figure}
\begin{tabular}{cc}
\includegraphics[width=0.53 \textwidth]{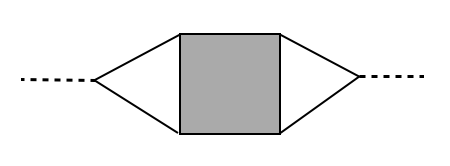}
\put(-129,39){\small $SU(\nc)$}
 &
\includegraphics[width=0.4 \textwidth]{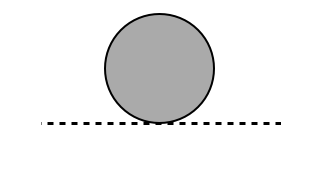}
\put(-105,56){\small $SU(\nc)$}
\\
\end{tabular}
\caption{Diagrams contributing to the two-point function of electromagnetic currents to leading order in the electromagnetic coupling $\alpha_\mt{EM}$. The external, dotted lines correspond to photons with momentum $k$, while the shaded blobs denote full resummations of the $SU(\nc)$ diagrams, to all orders in $\lambda$.}\label{diag}
\end{figure}

We do not know the gravitational dual of the full $SU(\nc)\times U(1)$ theory, but fortunately this will not be necessary for our purposes. It was in fact shown in \cite{CaronHuot:2006te} that to compute the two-point correlation function of the electromagnetic current (\ref{current}) to leading order in the electromagnetic coupling $\alpha_\mt{EM}$, it is enough to consider the $SU(\nc)$ theory only, whose dual is known from \cite{Mateos:2011ix,Mateos:2011tv}. Our computation will then be to leading order in $\alpha_\mt{EM}$, since the coupling of the photons to the surrounding medium is small, but fully non-perturbative in the 't~Hooft coupling $\lambda$ of the $SU(\nc)$ theory. This is depicted diagrammatically in Fig.~\ref{diag}, where the shaded blobs stand for the all-order resummations of the $SU(\nc)$ theory diagrams, while the external legs are the $U(1)$ theory photons.

In general, photon production in differential form is given by the expression \cite{lebellac,CaronHuot:2006te,Mateos:2007yp}
\bea
\frac{d\Gamma}{d\vec k} = \frac{e^2}{(2\pi)^3 2|\vec k|}\Phi(k)\sum_{s=1,2} \epsilon^\mu_{(s)}(\vec k)\,  \epsilon^\nu_{(s)}(\vec k)\, \chi_{\mu\nu}(k)\Big|_{k^0=|\vec k|}\,,
\label{diff}
\eea
with $k^\mu=(k^0,\vec k)$ the photon null momentum and $\Phi(k)$ the distribution function, which for thermal equilibrium, as in our case, reduces to the Bose-Einstein distribution $n_B(k^0)=1/(e^{k^0/T}-1)$. The spectral density is $\chi_{\mu\nu}(k)=-2 \mbox{ Im } G^\mt{R}_{\mu\nu}(k)$, with
\bea
G^\mt{R}_{\mu\nu}(k) = -i \int d^4x \, e^{-i k\cdot x}\, \Theta(t) \vev{[\jem_\mu(x),\jem_\nu(0)]}
\eea
the retarded correlator of two electro-magnetic currents $\jem_\mu$.

\begin{figure}
    \begin{center}
        \includegraphics[width=0.90\textwidth]{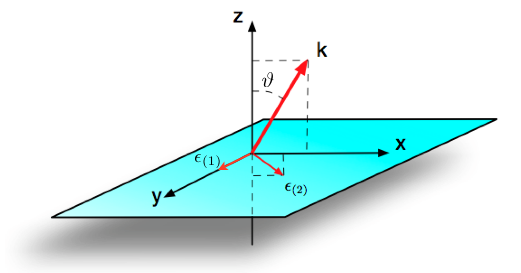}
        \caption{Photon momentum and polarization vectors. Because of the rotational symmetry in the $xy$-plane, the momentum can be chosen to be contained in the $xz$-plane, forming an angle $\vartheta$ with the $z$-direction. $\vec \epsilon_{(1)}$ is oriented along the $y$-direction and $\vec\epsilon_{(2)}$ is contained in the $xz$-plane, orthogonally to $\vec k$.}
        \label{momentum}
    \end{center}
\end{figure}

Each term of the sum in (\ref{diff}) stands for the number of photons emitted with polarization vector $\vec\epsilon_{(s)}$. The two polarization vectors are mutually orthogonal and orthogonal to $\vec k$. Given the $SO(2)$ symmetry in the $xy$-plane, we can choose without  loss of generality $\vec k$ to lie in the $xz$-plane, forming an angle $\vartheta$ with the $z$-direction -- see Fig.~\ref{momentum}. 
Specifically, we set
\be
\vec k = k_0 (\sin \vartheta, 0, \cos \vartheta) \,.
\ee
This means that we can choose the polarization vectors as 
\be
\vec\epsilon_{(1)}=(0,1,0)\sac \vec\epsilon_{(2)} = (\cos \vartheta, 0, - \sin \vartheta) \,.
\ee
Production of photons with polarization $\vec\epsilon_{(1)}$ is then proportional to $\chi_{yy} \sim \mbox{Im}\, \langle \jem_y \jem_y \rangle$, whereas for those with polarization $\vec\epsilon_{(2)}$ it is proportional to\footnote{Note that $\chi_{xz}=\chi_{zx}$; see e.g.~\cite{CaronHuot:2006te}.} 
\be
\epsilon^\mu_{(2)}\,  \epsilon^\nu_{(2)}\, \chi_{\mu\nu} = 
\cos^2 \vartheta \, \chi_{xx} + \sin^2 \vartheta \, \chi_{zz} 
- 2 \cos \vartheta \sin \vartheta  \, \chi_{xz} \,. 
\label{combination}
\ee
From this expression we see that we need to compute the three correlators 
\be
G^\mt{R}_{xx} \sim \langle \jem_x \jem_x \rangle  \sac G^\mt{R}_{zz} \sim \langle \jem_z \jem_z \rangle 
\sac G^\mt{R}_{xz} \sim \langle \jem_x \jem_z \rangle    
\ee
and then add them up according to \eqn{combination}. In the following section we will see how these correlators can be obtained from gravity.


\section{Gravity set-up}
\label{sec2}

The dual gravitational background for the theory (\ref{andef}) at finite temperature is the type IIB supergravity geometry found in \cite{Mateos:2011ix,Mateos:2011tv}, whose string frame metric reads
\bea
ds^2=\frac{L^2}{u^2}\left(-{\cal B} {\cal F} \, dt^2+dx^2+dy^2+{\cal H} dz^2 +\frac{du^2}{{\cal F}}\right)+L^2\, e^{\frac{1}{2}\phi}d\Omega_5^2\,,
\label{metric}
\eea
with ${\cal H}=e^{-\phi}$ and $\Omega_5$ the volume form of a round 5-sphere
\bea
d\Omega_5^2 = d\psi^2 + \cos^2\psi \, d\varphi^2+\sin^2\psi\,  d\Omega^2_3\,.
\eea
The gauge theory coordinates are $(t,x,y,z)$ while $u$ is the AdS radial coordinate, with the black hole horizon lying at $u=\uh$ (where $\cf$ vanishes) and the boundary at $u=0$. As mentioned already, we refer to the $z$-direction as the longitudinal direction and to $x$ and $y$ as the transverse directions. $L$ is the common radius of the first metric factor and of the sphere and is set to unity in the following. Besides the metric and the dilaton $\phi$, the forms
\bea
F_5 =4 \left(\Omega_5 + \star \Omega_5\right)\,,\qquad F_1=a \, dz
\eea
are also turned on, with $a$ being a parameter with units of energy that controls the degree of anisotropy of the system. The potential for the 1-form is a linear axion $\chi=a\, z$. This acts as an isotropy-breaking external source that forces the system into an anisotropic equilibrium state.

\begin{figure}[tb]
\begin{center}
\begin{tabular}{cc}
\includegraphics[scale=0.8]{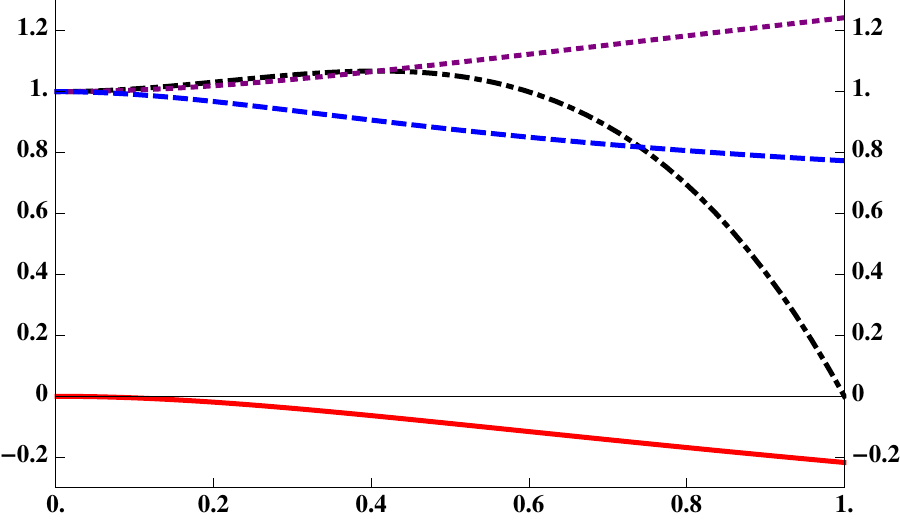}
\put(-109,-10){\small $u/\uh$}
\put(-109,-14){$$}
\put(-150,17){$\phi$}
\put(-70,115){$\ch$}
\put(-30,88){$\cb$}
\put(-43,45){$\cf$}
&
\includegraphics[scale=0.8]{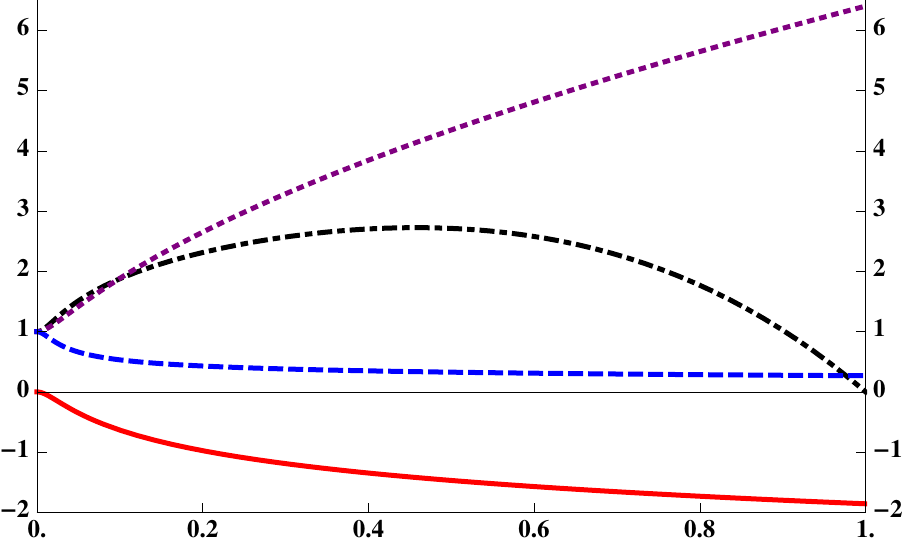}
\put(-109,-10){\small $u/\uh$}
\put(-109,-14){$$}
\put(-55,17){$\phi$}
\put(-70,113){$\ch$}
\put(-150,45){$\cb$}
\put(-53,67){$\cf$}
\end{tabular}
\caption{\small Metric functions for $a/T\simeq 4.4$ (left) and $a/T\simeq 86$ (right).
\label{plots}}
\end{center} 
\end{figure}
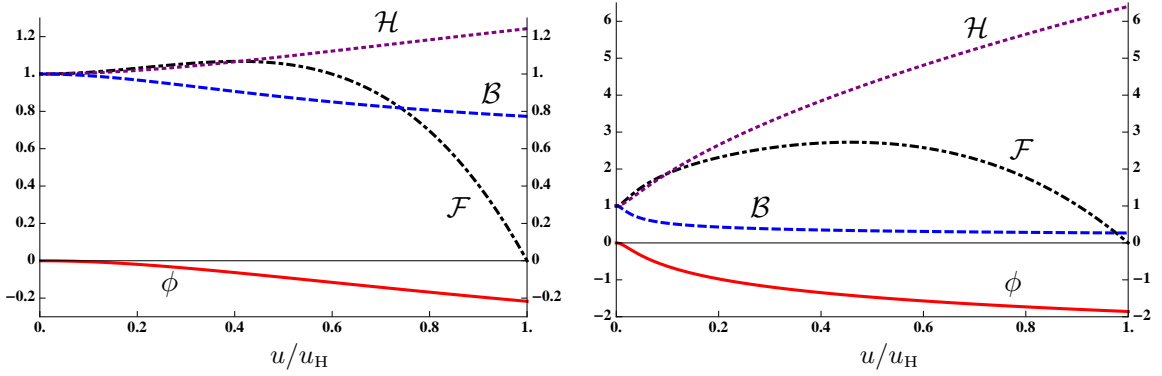

\begin{figure}
\begin{center}
\includegraphics[scale=.8]{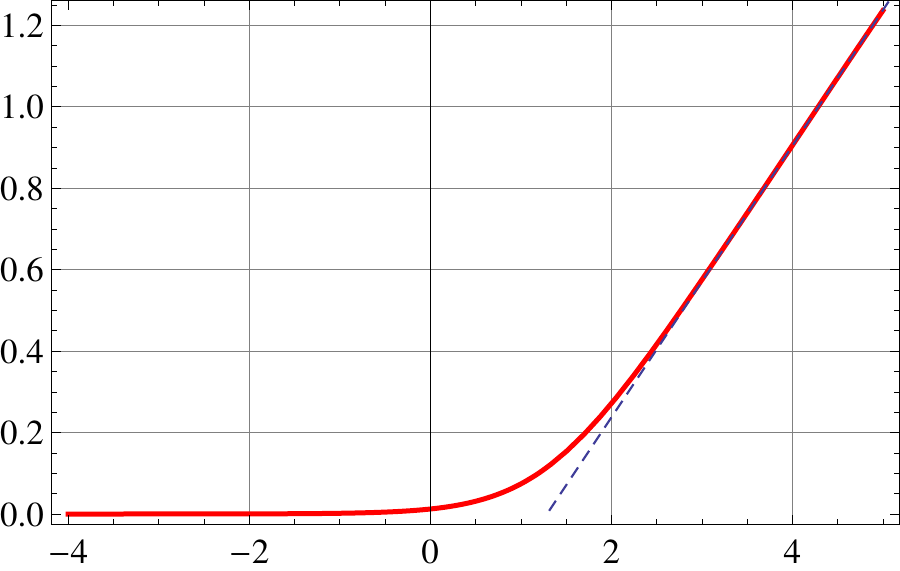}
\put(-50,-10){\small $\log (a/T)$}
\put(-230,70){\rotatebox{90}{\small $\log (s/ s_\mt{iso})$}}
\caption{\small Log-log plot of the entropy density as a function of $a/T$. The dashed blue line is a straight line with slope 1/3.}
\label{scalings}
\end{center} 
\end{figure}

The functions ${\cal B}, {\cal F}$, and $\phi$ depend solely on $u$; they are known analytically in limiting regimes of low and high temperature, and numerically in intermediate regimes \cite{Mateos:2011tv}. Representatives of these functions for two values of $a/T$ are plotted in Fig.~\ref{plots}. 
For $u\to 0$ (independently of the value of $a$) they asymptote to the $AdS_5\times S^5$ metric, $\cf=\cb=\ch=1$ and $\phi=0$, while for $a=0$ they reduce to the black D3-brane solution
\be
\cb=\ch=1\,, \qquad \phi=\chi=0\,, \qquad \cf=1-\frac{u^4}{\uh^4}\,,
\label{isometric}
\ee
which has temperature  and entropy density given by \cite{peet}
\be
T_\mt{iso}=\frac{1}{\pi\uh}\,, \qquad s_\mt{iso}= \frac{\pi^2}{2} \nc^2 T^3 \,.
\label{siso}
\ee
The temperature and entropy density of the anisotropic geometry are given by \cite{Mateos:2011tv}
\be
T=\frac{e^{-\frac{1}{2}\phi_\mt{H}}\sqrt{{\cal B}_\mt{H}}(16+a^2 \uh^2 e^{\frac{7}{2}\phi_\mt{H}})}{16 \pi \uh}\,,
\qquad
s=\frac{\nc^2}{2\pi\uh^3}e^{-\frac{5}{4}\phi_\mt{H}}\,,
\label{temperature}
\ee
where $\phi_\mt{H}=\phi(u=\uh)$ and $\cb_\mt{H}=\cb(u=\uh)$. As depicted in Fig.~\ref{scalings}, the entropy density of the system interpolates smoothly between the isotropic scaling above for small $a/T$ and the scaling \cite{Mateos:2011tv,ALT}
\be
s \simeq 3.21\, \nc^2 T^3 \left(\frac{a}{T}\right)^\frac{1}{3} \,,
\label{saniso}
\ee
for large $a/T$, the transition between the two behaviors taking place at approximately $a/T\simeq 3.7$. The space can then be interpreted as a domain-wall-like solution interpolating between an AdS geometry in the UV and a Lifshitz-like geometry in the IR, with the radial position at which the transition takes place being set by the anisotropic scale $a$: when $T\gg a$ the horizon lies in the asymptotic AdS region with scaling (\ref{siso}), whereas for $T\ll a$ it lies in the anisotropic region with scaling (\ref{saniso}).

A feature of the anisotropic geometry of \cite{Mateos:2011ix,Mateos:2011tv} is the presence of a conformal anomaly that appears during the renormalization of the theory, introducing a reference scale $\mu$. This anomaly implies that some physical quantities (such as, for example, the energy density and pressures) do not depend only on the ratio $a/T$, but on two independent dimensionless ratios that can be built out of $a$, $T$, and $\mu$.\footnote{The metric functions do depend solely on $a/T$, which is why we have used this ratio to distinguish between two representative examples in Fig.~\ref{plots}.} Fortunately, as we shall see in the following, all the quantities computed in this paper are not affected by this anomaly and will be independent of $\mu$. In particular,  they will be functions only of two independent dimensionless ratios that can be built out of $a$, $T$, and $k_0$ (e.g. $a/T$ and $k_0/T$). 


\subsection{Flavor branes}

The introduction of $N_\mt{f}$ flavors of massless quarks is achieved by placing $N_\mt{f}$ probe D7-branes in the background (\ref{metric}). Given that the quarks are massless, any temperature different from zero in the background will imply that these quarks are in a deconfined phase (the `black-hole embedding' \cite{thermobrane}) and also that the induced metric over the D7's will be the metric (\ref{metric}) with a three-sphere instead of a five-sphere. The complete system can then be thought of as a D3/D7 system with two different kinds of D7-branes, one kind sourcing the anisotropy \cite{Mateos:2011tv,ALT}%
\footnote{These branes are smeared homogeneously along the $z$-direction and can be thought of as giving rise to a density $n_\mt{D7}=N_\mt{D7}/L_z$ of extended charges. This charge density is related to the anisotropy parameter $a$ through $a=\gym^2 n_\mt{D7}/4\pi$ \cite{Mateos:2011tv}.
\label{footnote}} 
and the other kind sourcing flavor \cite{flavor1,flavor2}:
\bea
\begin{array}{l| cccc|c|ccc}
& t & x & y & z & u & \psi & \varphi & \Omega_3 \\
\hline 
N_\mt{c} ~~ \mbox{  D3 } & \times & \times & \times & \times & & & & \\
N_\mt{D7} ~ \mbox{  D7 } &  \times &  \times & \times & & &  \times &  \times &  \times \\
N_\mt{f}~ ~~\mbox{ D7 } &  \times &  \times &  \times & \times & \times &  & & \times 
\end{array}\,.
\eea
Throughout our calculations, the value of $N_\mt{D7}$ (and consequently of $a$, see footnote~\ref{footnote}) can be arbitrary, but we will assume that flavor is quenched, $N_\mt{f}\ll N_\mt{c}$. As  discussed in \cite{Mateos:2011tv}, the full backreaction of the `anisotropic' D7-branes is already included in the geometry (\ref{metric}), through the presence of the axion field $\chi$. In the following, when talking about D7-branes we will refer exclusively to the flavor D7-branes, which are probes in the fixed background (\ref{metric}). 

As argued in \cite{CaronHuot:2006te}, to study photon emission to leading order in $e$ it suffices to evaluate the correlators needed for (\ref{diff}) in the $SU(\nc)$ gauge theory with no dynamical photons. At strong 't Hooft coupling and large $\nc$, these correlators can be calculated holographically. Indeed, global symmetries of the gauge theory are in one-to-one correspondence with gauge symmetries on the gravity side, and each conserved current of the gauge theory is dual to a gauge field on the gravity side. 

Let $A_m$ $(m=0, \ldots, 7)$ be the gauge field associated to the overall $U(1)\subset U(\nf)$ gauge symmetry on the D7-branes. Upon dimensional reduction on the $3$-sphere wrapped by the D7-branes, $A_m$ gives rise to a massless gauge field  $(A_\mu, A_u)$, three massless scalars, and a tower of massive Kaluza-Klein (KK) modes. All these fields propagate on the five non-compact dimensions of the D7-branes. We will work in the gauge $A_u=0$,\footnote{This gauge choice will be immaterial in the following, since we shall switch to gauge invariant quantities, but it has the advantage of simplifying our formulas.} and we will consistently set the scalars and the higher KK modes to zero, since these are not of interest here. The gauge field $A_\mu$ is the desired dual to the conserved electromagnetic current $\jem_\mu$ of the gauge theory. According to the prescription of \cite{duality2,duality3}, correlation functions of $\jem_\mu$ can be calculated by varying the string partition function with respect to the value of $A_\mu$ at the boundary of the spacetime \eqn{metric}. 

We now proceed to write down the action for the D7-branes. It is easy to realize that there is no Wess-Zumino coupling of the branes to the background $F_5$, because of the particular brane orientation that has been chosen. A priori there could be a coupling to the background axion
\bea
\int_\mt{D7} \hskip -.1cm \chi \,  e^{2\pi \ell_s^2  F}\,, 
\eea
but this would be quartic in the $U(1)$ field strength $F=dA$ and therefore not relevant for the present computation, where we only need 2-point functions.

This means that the Dirac-Born-Infeld (DBI) action is all we need to consider:
\bea
S&=&-\nf \, T_\mt{D7} \int_\mt{D7} \hskip -.2cm d^8\sigma \, e^{-\phi} \sqrt{-\det\left(g+2\pi \ell_s^2  F\right)}\,,
\label{DBI}
\eea
where $g$ is the induced metric on the D7-branes and $T_\mt{D7} = 1/(2\pi \ell_s)^7 g_s \ell_s$ is the D7-brane tension. To obtain the equations of motion for $A_\mu$, it suffices to expand the action above and use the quadratic part only:
\be
S = - \nf T_\mt{D7} \int_\mt{D7} d^{8}\sigma \, e^{-\phi} \sqrt{-\det g} 
 \, \frac{(2\pi \ell_s^2)^2}{4} F^2  \,,
\label{sdq}
\ee
where $F^2 = F_{mn} F^{mn}$. The embedding of the branes inside the $S^5$ of the geometry can be parametrized by an angle $\cos\vartheta\equiv \psi(u)$ and the induced metric on the branes is given by
\bea
ds^2_\mt{D7}&=& \frac{1}{u^2}\left(-\cf \cb\,   dt^2+dx^2+dy^2+\ch\,  dz^2\right)+
\frac{1-\psi^2+u^2 \cf e^{\frac{1}{2}\phi} \psi'^2}{u^2 \cf(1-\psi^2)}du^2 
\cr && 
+e^{\frac{1}{2}\phi}(1-\psi^2)d\Omega_3^2\,.
\eea
The case of massless quarks corresponds to the equatorial embedding of the D7-branes, namely to $\psi=0$. After the dimensional reduction on the three-sphere, the action reduces to
\be
S =-K_\mt{D7} \int dt\, d\vec x \, du\frac{ e^{-\frac{3}{4}\phi}\sqrt{\cb}}{u^5} F^2 \,,
\label{sq5}
\ee
where
\be
K_\mt{D7}=2 \pi^4  N_\mt{f}T_\mt{D7} \ell_s^4=\frac{1}{16\pi^2}\nc\nf\,,
\ee
and $F_m$ is restricted to the components  $m=(\mu,u)$.

As argued in \cite{thermobrane,Mateos:2007yp}, in order to calculate the photon emission rate, we may consistently proceed by assuming the equatorial embedding of the D7-branes to be fixed in the absence of the gauge field, and then solving for the gauge field on that embedding. No modes of the metric or of the background fields will be excited. As mentioned above, we set the components of the gauge field on the three-sphere wrapped by the D7-branes to zero and we Fourier decompose the remaining components as
\be
A_\mu(t, \vec x, u) = \int \frac{d k^0 d \vec k}{(2\pi)^{4}} \, 
e^{-i k^0 t + i \vec k \cdot \vec x} \, A_\mu (k^0, \vec k, u) \,,\qquad \vec k=k_0 (\sin \vartheta, 0, \cos \vartheta)\,.
\label{fourier}
\ee
This is possible because the state we consider, although anisotropic, is translationally invariant along the gauge theory directions \cite{Mateos:2011tv}. Doing so, the equations for the gauge field deriving from (\ref{sq5}) split into the following single, decoupled equation for $A_y$ (primes denote derivatives with respect to $u$):
\bea
\left(M g^{uu}g^{yy}A'_y\right)'-M g^{yy}\left(g^{tt}k_0^2+g^{xx}k_x^2+g^{zz}k_z^2\right)A_y=0\,,
\label{eomy}
\eea
together with a coupled system of three equations for the  remaining components $A_{t,x,z}$:
\begin{eqnarray}
&& \hskip -.7cm  
(M g^{uu}g^{tt}A'_t)'-M g^{tt}\left[ g^{xx}k_x(k_xA_t-k_0A_x)+g^{zz}k_z(k_zA_t-k_0A_z)\right] = 0\,, 
\label{eom1} \\
&& \hskip -.7cm 
(M g^{uu}g^{xx}A'_x)'-M g^{xx}\left[ g^{tt}k_0(k_0A_x-k_xA_t)+g^{zz}k_z(k_zA_x-k_xA_z)\right] = 0\,, 
\label{eom2} \\
&& \hskip -.7cm 
(M g^{uu}g^{zz}A'_z)'-M g^{zz}\left[ g^{tt}k_0(k_0A_z-k_zA_t)+g^{xx}k_x(k_xA_z-k_zA_x)\right] = 0\, . 
\label{eom3}
\end{eqnarray}
The inverse metric can be read off directly from (\ref{metric}) and $M$ is the factor appearing in the action (\ref{sq5}):
\bea
M\equiv \frac{e^{-\frac{3}{4}\phi}\sqrt{{\cal B}}}{u^5}\,.
\eea
Equations (\ref{eomy}) to (\ref{eom3}) constitute the set of equations that we shall solve in the next section, with the appropriate boundary conditions, to obtain the correlation functions of the electromagnetic currents $\jem_\mu$.


\section{Photon production from holography}
\label{sec3}

The Lorentzian AdS/CFT prescription to compute the correlation functions of the electromagnetic current is given in \cite{recipe} (see also, e.g., \cite{PSS,PSS2,KS}). We need to isolate the terms with two radial derivatives in the action (\ref{sq5}), write the latter as a boundary term by integration by parts and compute it on-shell. Varying with respect to the values of $A_\mu$ at $u=0$ (which correspond to the boundary theory currents) gives the desired correlation functions. 

We start by writing the boundary action  as
\be 
S_\epsilon=
-2K_\mt{D7}\int dt\, d\vec{x}\, \left[\frac{e^{- \frac{3}{4} \phi}\sqrt{{\cal B}}{\cal F}}{u}
\left( -\frac{1}{{\cal B}{\cal F}}A_tA_t'+A_xA_x'+A_yA_y'+e^\phi A_zA_z'\right)\right]_{u=\epsilon}\,, 
\label{boundactexp}
\ee
where the limit $\epsilon\rightarrow 0$ will be taken.

From this boundary action and the equations of motion (\ref{eomy})-(\ref{eom3}), we see that we can carry out the calculation for the spectral density $\chi_{yy}$ independently of all the others; this calculation is very similar to that in \cite{Mateos:2007yp}. The remaining spectral densities in (\ref{combination}) are more involved and will require a strategy that was originally developed in \cite{Kaminski:2009dh} to deal with operator mixing.


\subsection{Isotropic limit}

In order to compare our results with the photon production rate in an isotropic plasma, we will need expressions for the isotropic correlators both as a function of the temperature of the plasma and as a function of the entropy density of the plasma.  

We start then by considering the isotropic limit, which is obtained by plugging (\ref{isometric}) in the equations above. We may also set $\vartheta=\pi/2$, since in the isotropic system we have an $SO(3)$ symmetry that allows us to align the photon momentum along a particular direction, say the $x$-direction. The equation to solve is obtained from (\ref{eomy}) and reads\footnote{This equation is exactly equivalent to eq. (4.18) of \cite{Mateos:2007yp} under the change of coordinates $u_\mt{here}=\uh\sqrt{2 u_\mt{there}}$.}
\be
0=u\left(1-\frac{u^4}{\uh^4}\right)^2A_\mt{iso}''-\left(1-\frac{u^4}{\uh^4}\right)\left(1+3 \frac{u^4}{\uh^4}\right)A_\mt{iso}'+k_0^2\frac{u^5}{\uh^4} A_\mt{iso}
\,.
\ee
The solution of this equation that is infalling at the horizon reads
\be
A_\mt{iso}=\left(1-\frac{u^2}{\uh^2}\right)^{-i\frac{\wn}{2}}\left(1+\frac{u^2}{\uh^2}\right)^{-\frac{\wn}{2}}
\hskip-.2cm {}_2F_1\left(1-\frac{1+i}{2}\wn,\, -\frac{1+i}{2}\wn,\, 1-i\wn;\, \frac{\uh^2-u^2}{2\uh^2}\right)\,,
\ee
where
\be
\wn=\frac{k_0}{2\pi T}=\frac{k_0\uh}{2}
\ee
is the dimensionless frequency as customary in the quasi-normal mode literature. 

\begin{figure}[h!]
    \begin{center}
        \includegraphics[width=0.5\textwidth]{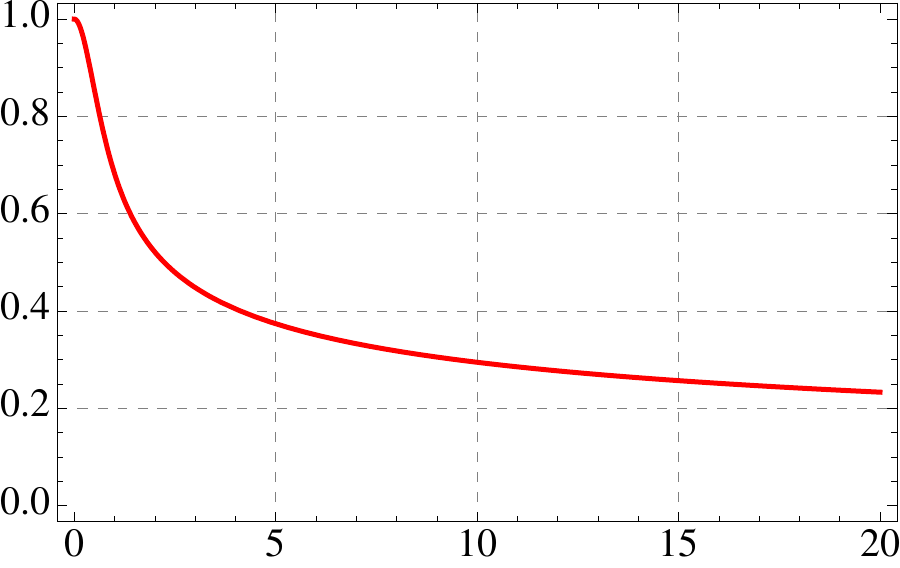}
         \put(-240,47){\rotatebox{90}{$\chi_\mt{iso}/8\tilde{\cal N}_\mt{D7}\,\wn$}}
         \put(10,5){$\wn$}
        \caption{The isotropic correlator $\chi_\mt{iso}$. }
        \label{plotiso}
    \end{center}
\end{figure}

Rotational invariance implies that the sum over polarizations in (\ref{diff}) be equal to the following expression 
\be
\chi_\mt{iso}=-2\, \mbox{Im}\,\left(G^\mt{R}_{yy}+G^\mt{R}_{zz}\right)_\mt{iso}=-4\, \mbox{Im}\, G^\mt{R}_\mt{iso}\,,
\label{spectralisotropic}
\ee
which can be computed from the prescription in \cite{recipe} using
\be
G_\mt{iso}^\mt{R} = -\frac{4 K_\mt{D7}}{|A_\mt{iso}(\wn, 0)|^2}  \lim_{u \rightarrow 0}  
 \left(1-\frac{u^4}{\uh^4}\right)\frac{1}{u}A_\mt{iso}^*(\wn,u) A'_\mt{iso}(\wn, u) \,.
\ee
The imaginary part of this expression turns out to be independent of $u$, and is actually easier to evaluate the limit at the horizon rather than at the boundary. The result of the limit is
\be
\lim_{u \rightarrow \uh}  
 \left(1-\frac{u^4}{\uh^4}\right)\frac{1}{u}A_\mt{iso}^*(\wn,u) A'_\mt{iso}(\wn, u)=i\frac{2\wn}{2^\wn \uh^2}\,.
\ee
Putting everything together, one finally obtains that the spectral density is given by~\cite{Mateos:2007yp}
\be
\chi_\mt{iso}= \frac{8\,  \tilde{\cal N}_\mt{D7}\, \wn }{2^\wn\, 
\left|{}_2F_1\left(1-\frac{1+i}{2}\wn,\, -\frac{1+i}{2}\wn,\, 1-i\wn;\, \frac{1}{2}\right)\right|^2}
\,,
\ee
where
\be
\tilde{\cal N}_\mt{D7}=4\frac{K_\mt{D7}}{\uh^2}=\frac{1}{4}\nc\nf T^2\,.
\ee
The dimensionless combination $\chi_\mt{iso}/8 \tilde{\cal N}_\mt{D7} \wn $ is plotted in Fig.~\ref{plotiso}.
We can now express $\tilde{\cal N}_\mt{D7}$ and $\wn$ in the formula above either in terms of the temperature or in terms of the entropy density of the isotropic plasma, using (\ref{siso}). We obtain
\be
\chi_\mt{iso}(T)= \frac{\nc\nf\, T\, k_0 }{2^\frac{k_0}{2\pi T} \pi \, 
\left|{}_2F_1\left(1-\frac{1+i}{4\pi T}k_0,\, -\frac{1+i}{4\pi T}k_0,\, 1-i\frac{k_0}{2\pi T};\, \frac{1}{2}\right)\right|^2}\,,
\label{chiisoT}
\ee
and
\be
\chi_\mt{iso}(s)= \frac{\nc^\frac{1}{3}\nf\, s^\frac{1}{3}\, k_0 }{2^{\wn_s-\frac{1}{3}}\pi^\frac{5}{3} \, 
\left|{}_2F_1\left(1-\frac{1+i}{2}\wn_s,\, -\frac{1+i}{2}\wn_s,\, 1-i\wn_s;\, \frac{1}{2}\right)\right|^2}
\,,\qquad 
\wn_s=\frac{\nc^\frac{2}{3}k_0}{2^\frac{4}{3}\pi^\frac{1}{3}s^\frac{1}{3}}\,.
\label{chiisos}
\ee
In the following sections we shall use these expressions to normalize the anisotropic results. The two different normalizations will allow us to compare the anisotropic photon production either with an isotropic plasma at the same temperature but different entropy density or with an isotropic plasma at the same entropy density but different temperature.


\subsection{Spectral density for the polarization $\epsilon_{(1)}$}

After the isotropic case warm-up, we proceed to compute $\chi_{yy}$. This can be obtained independently of all the other spectral densities, given that (\ref{eomy}) is not coupled to the remaining equations of motion and that there is no mixing term in the action between $A_y$ and the other gauge fields. The corresponding correlation function is
\be
G_{yy}^\mt{R} = - \frac{4K_\mt{D7}}{\left|A_y(k_0, 0)\right|^2}  \lim_{u \rightarrow 0}  
Q(u) A_y^*(k_0,u) A'_y(k_0, u) \,,\label{cyy}
\ee
where 
\bea
Q(u)\equiv \frac{e^{-\frac{3}{4}\phi}\sqrt{{\cal B}}{\cal F}}{u}   \,.
\label{Q}
\eea
The spectral density reads
\be
\chi_{(1)}\equiv \chi_{yy}=\frac{\nc\nf}{2\pi^2\left|A_y(k_0, 0)\right|^2}\,\mbox{Im}\,  
\lim_{u \rightarrow 0}  Q(u) A_y^*(k_0,u) A'_y(k_0, u) \,.
\ee

In the anisotropic case the metric components appearing in the equations of motion are only known numerically for generic values of $a/T$. This means that we have to resort to numerics in order to integrate (\ref{eomy}) and obtain the solution for $A_y$. To impose the appropriate boundary conditions at the horizon we use the near-horizon expansion of the metric given in \cite{Mateos:2011tv} and expand (\ref{eomy}). We see that close to the horizon $A_y$ behaves like $(u-\uh)^\nu a_y(u)$, where $a_y(u)$ is some function regular at $\uh$. The exponent $\nu$ turns out to be given by the usual result 
\be
\nu=\pm i \frac{\wn}{2}\, , \qquad \wn=\frac{k_0}{2\pi T}\,,
\ee 
where the temperature appearing in this formula is now the anisotropic one, given by (\ref{temperature}). To impose the infalling wave condition at the horizon and therefore get the retarded correlator, we choose the negative sign for $\nu$. Once the infalling solution has been selected, the expansion is unique up to an overall multiplicative factor.

\begin{figure}
\begin{center}
\begin{tabular}{cc}
\setlength{\unitlength}{1cm}
\hspace{-0.9cm}
\includegraphics[width=7cm]{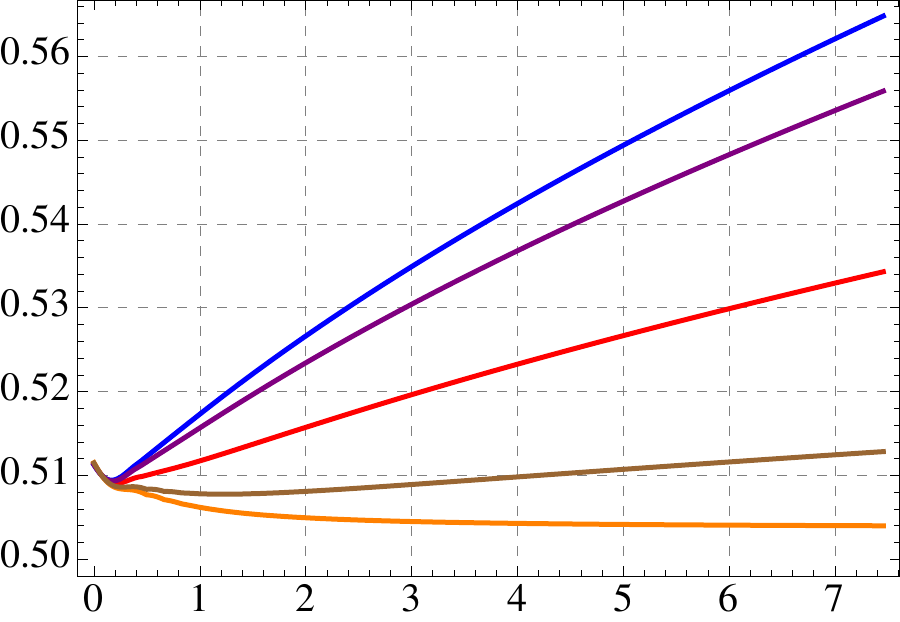} 
\qquad\qquad & 
\includegraphics[width=7cm]{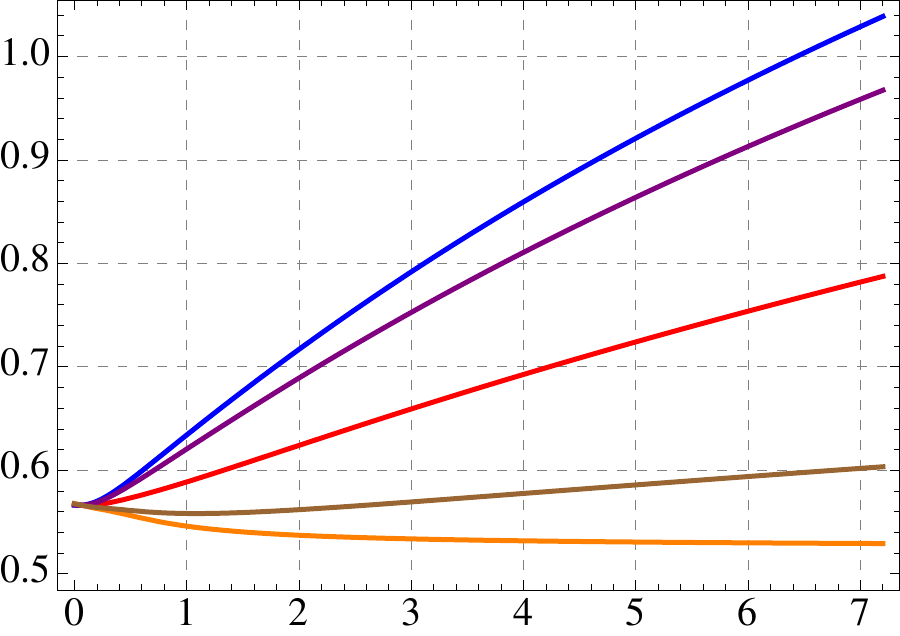}
\qquad
  \put(-455,40){\rotatebox{90}{$\chi_{(1)}/\chi_\mt{iso}(T)$}}
         \put(-250,-10){$\wn$}
         \put(-215,40){\rotatebox{90}{$\chi_{(1)}/\chi_\mt{iso}(T)$}}
         \put(-17,-10){$\wn$}
\\
(a) & (b)\\
& \\
\hspace{-0.9cm}
\includegraphics[width=7cm]{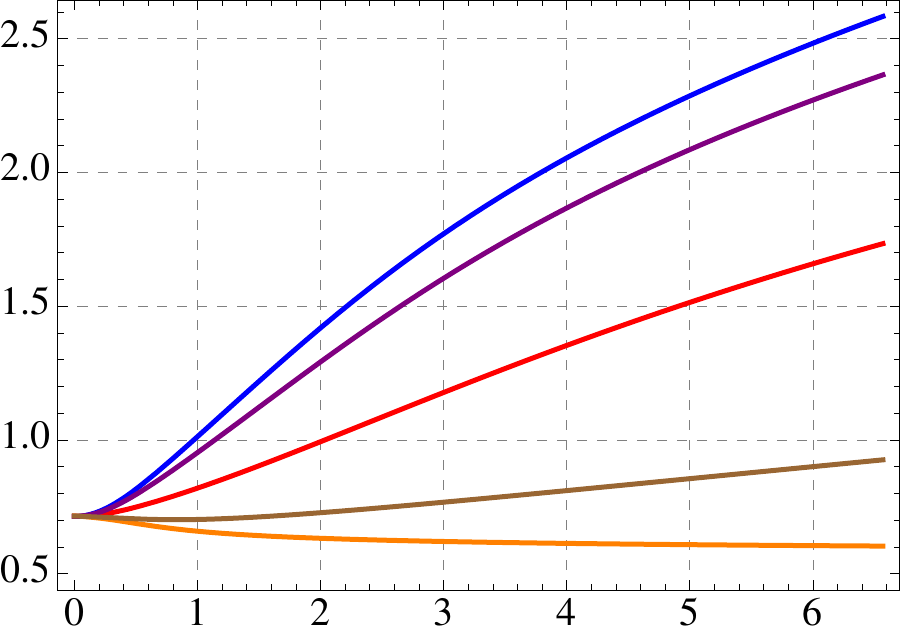} 
\qquad\qquad & 
\includegraphics[width=7cm]{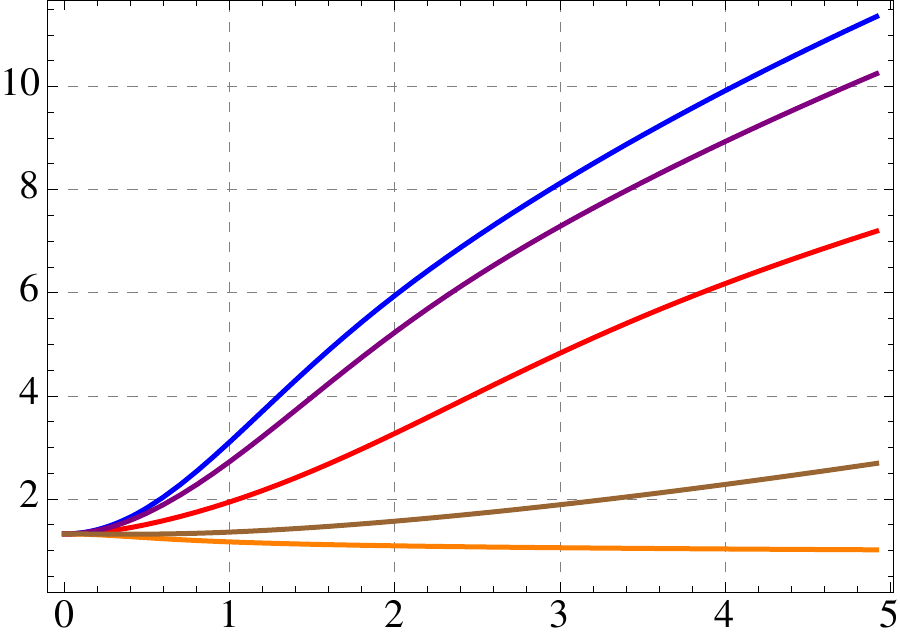}
\qquad
 \put(-455,40){\rotatebox{90}{$\chi_{(1)}/\chi_\mt{iso}(T)$}}
         \put(-250,-10){$\wn$}
         \put(-215,40){\rotatebox{90}{$\chi_{(1)}/\chi_\mt{iso}(T)$}}
         \put(-17,-10){$\wn$}
         \\
(c)& (d) 
\end{tabular}
\end{center}
\caption{\small Plots of the spectral density $\chi_{(1)}$ corresponding to the polarization $\epsilon_{(1)}$ normalized with respect to the isotropic result at fixed temperature $\chi_\mt{iso}(T)$. The curves correspond from top to bottom to the angles $\vartheta=0,\, \pi/8,\, \pi/4,\, 3\pi/8,\, \pi/2$. The four plots correspond to the cases $a/T=1.38$ (a), $4.41$ (b), $12.2$ (c), $86$ (d).}
\label{cyyplotT}
\end{figure}

\begin{figure}
\begin{center}
\begin{tabular}{cc}
\setlength{\unitlength}{1cm}
\hspace{-0.9cm}
\includegraphics[width=7cm]{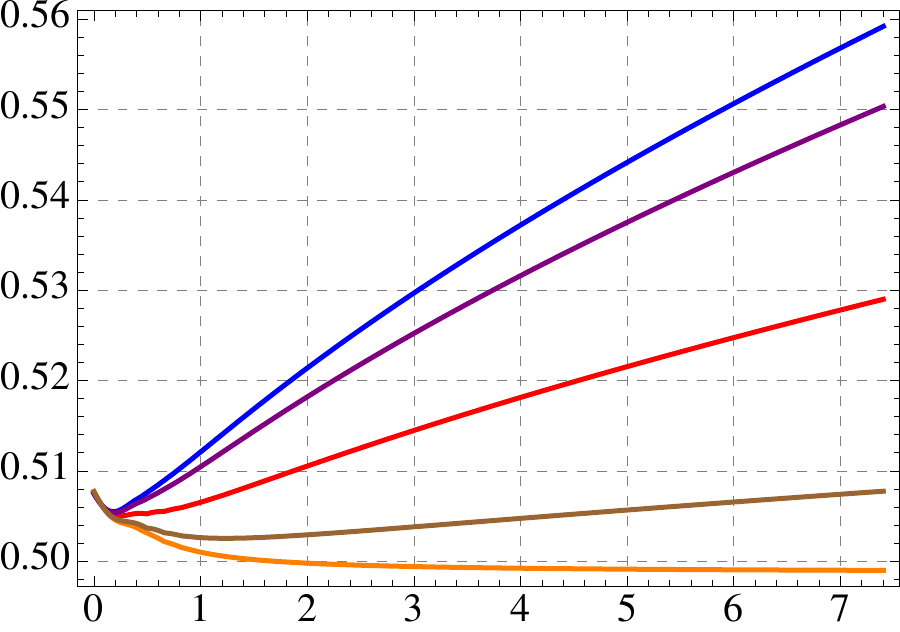} 
\qquad\qquad & 
\includegraphics[width=7cm]{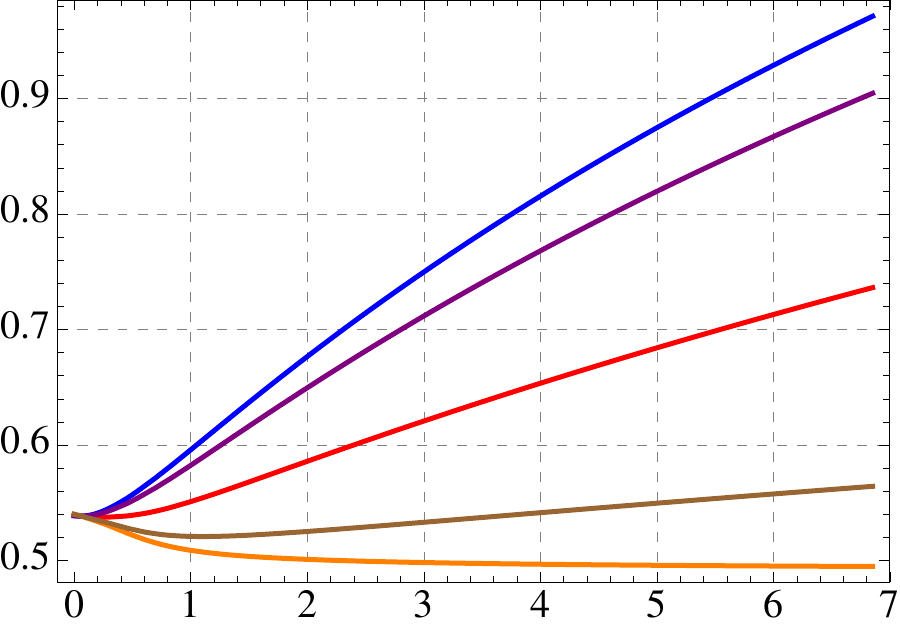}
\qquad
  \put(-455,40){\rotatebox{90}{$\chi_{(1)}/\chi_\mt{iso}(s)$}}
         \put(-250,-10){$\wn_s$}
         \put(-215,40){\rotatebox{90}{$\chi_{(1)}/\chi_\mt{iso}(s)$}}
         \put(-17,-10){$\wn_s$}
\\
(a) & (b)\\
& \\
\hspace{-0.9cm}
\includegraphics[width=7cm]{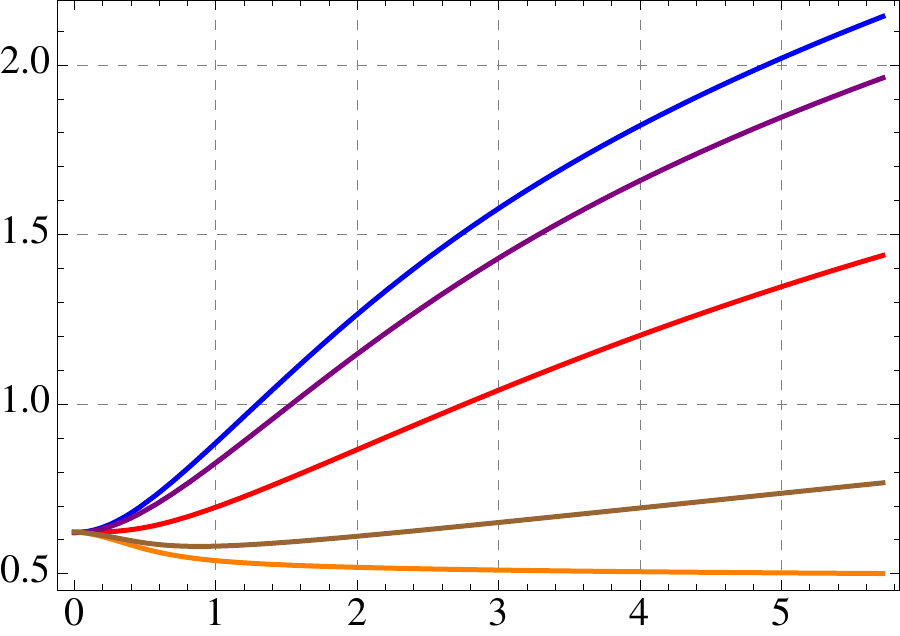} 
\qquad\qquad & 
\includegraphics[width=7cm]{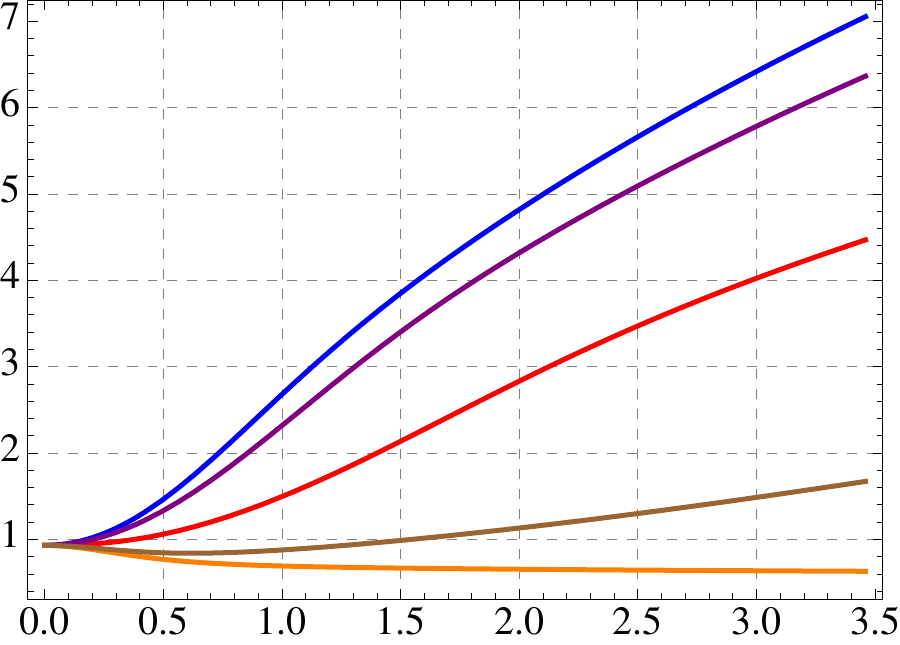}
\qquad
 \put(-455,40){\rotatebox{90}{$\chi_{(1)}/\chi_\mt{iso}(s)$}}
         \put(-250,-10){$\wn_s$}
         \put(-212,40){\rotatebox{90}{$\chi_{(1)}/\chi_\mt{iso}(s)$}}
         \put(-17,-10){$\wn_s$}
         \\
(c)& (d) 
\end{tabular}
\end{center}
\caption{\small Plots of the spectral density $\chi_{(1)}$ corresponding to the polarization $\epsilon_{(1)}$ normalized with respect to the isotropic result at fixed entropy density $\chi_\mt{iso}(s)$. The curves correspond from top to bottom to the angles $\vartheta=0,\, \pi/8,\, \pi/4,\, 3\pi/4,\, \pi/2$. The four plots correspond to the cases $a\nc^{2/3}/s^{1/3}=0.80$ (a), $2.47$ (b), $6.24$ (c), $35.5$ (d).}
\label{cyyplots}
\end{figure}

Now we carry out the numerical integration of (\ref{eomy}) for a number of values of the anisotropy and of the angle $\vartheta$ formed by the momentum $\vec{k}$  and the $z$-axis. The spectral density $\chi_{yy}$ is plotted in Figs.~\ref{cyyplotT} and \ref{cyyplots}, where it is compared with an isotropic plasma at the same temperature or at the same entropy density. Note that $\chi_{yy}$ in the isotropic limit goes to half of $\chi_\mt{iso}$, since the latter contains the (equal) contribution from $\chi_{zz}$ as well; see (\ref{spectralisotropic}).

Note also that at large $k_0$ (more specifically, for $k_0\gg a$) the spectral density does not converge to the isotropic limit (the horizontal line at $1/2$). This is explained by recalling that our results depend on two independent dimensionless ratios. Sending $k_0\to \infty$ while keeping $a$ and $T$ fixed is not the same as sending $a\to 0$ with $k_0$ and $T$ fixed, since the ratio $a/T$ remains finite in the first limit. 

\begin{figure}
\begin{center}
\begin{tabular}{cc}
\setlength{\unitlength}{1cm}
\hspace{-0.9cm}
\includegraphics[width=7cm]{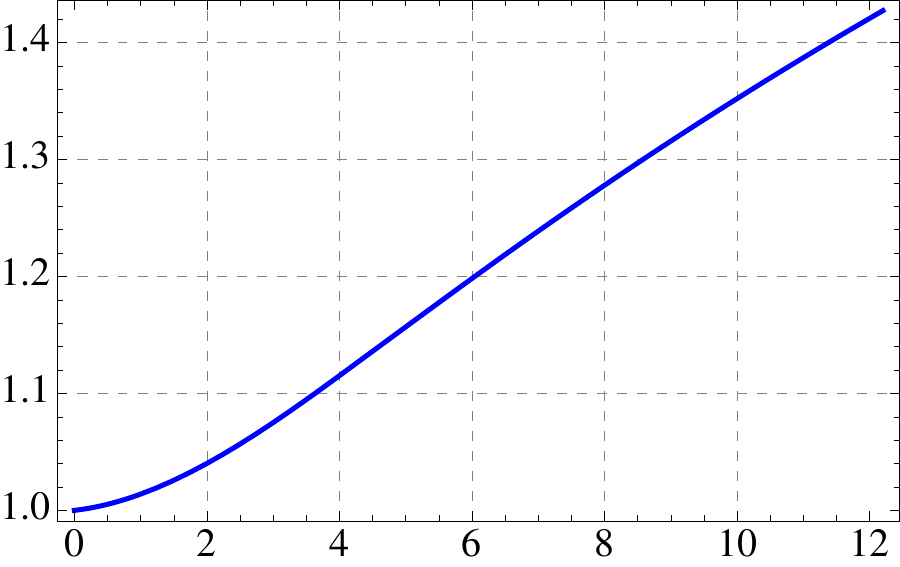} 
\qquad\qquad 
         & 
\includegraphics[width=7cm]{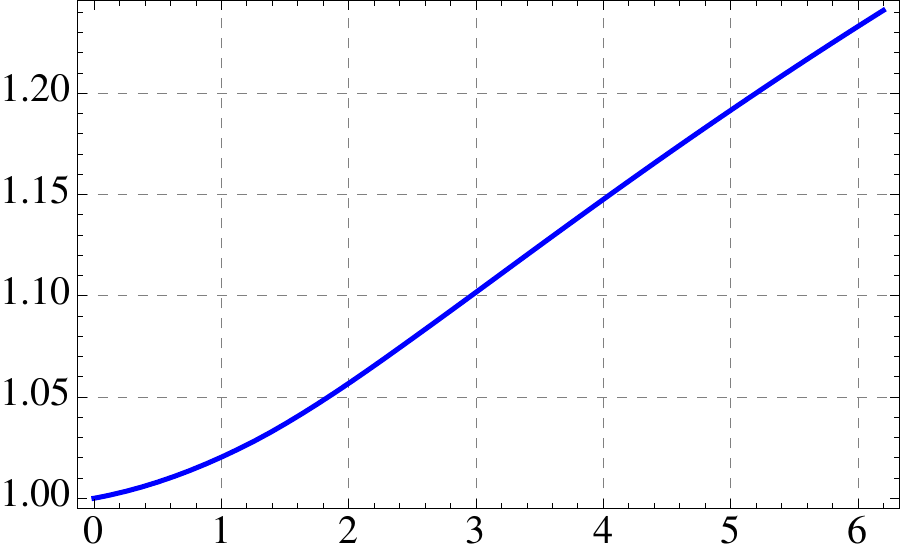}
\qquad
          \put(-455,80){\rotatebox{90}{$\sigma_{(1)}(T)$}}
         \put(-260,-15){$a/T$}
         \put(-217,80){\rotatebox{90}{$\sigma_{(1)}(s)$}}
         \put(-55,-15){$a\nc^{2/3}/s^{1/3}$}
         \\
         \end{tabular}
\end{center}
 \caption{\small Plot of the conductivity $\sigma_{(1)}$ corresponding to the polarization $\epsilon_{(1)}$ as a function of $a/T$ (left) and of $a\nc^{2/3}/s^{1/3}$ (right).
 }
\label{conductivity}
\end{figure}

The zero-frequency limit of the spectral density gives the electric DC conductivity. For photons with polarization $\epsilon_{(1)}$ this would be the conductivity along the transverse $y$-direction. The quantities
\bea
\sigma_{(1)}(T)&=&\lim_{k_0\to 0}\frac{\chi_{(1)}}{\chi_{(1),\mt{iso}}(T)}=\lim_{k_0\to 0}2\frac{ \chi_{(1)}}{\chi_\mt{iso}(T)}\,,\cr
\sigma_{(1)}(s)&=&\lim_{k_0\to 0}\frac{\chi_{(1)}}{\chi_{(1),\mt{iso}}(s)}=\lim_{k_0\to 0}2\frac{\chi_{(1)}}{\chi_\mt{iso}(s)}
\label{y_conductivity}
\eea
as functions of $a/T$ and of $a\nc^{2/3}/s^{1/3}$ are reported in Fig.~\ref{conductivity}. The plot on the left of that figure coincides with the perpendicular conductivity $\sigma_\perp$ obtained in \cite{rebhan_viscosity}. Note that the conductivity is to be normalized with respect to the isotropic result with the same polarization, i.e. $\lim_{a\to 0}\chi_{(1)}\equiv \chi_{(1),\mt{iso}}$ in the present case. 


\subsection{Spectral density for the polarization $\epsilon_{(2)}$}

Having found the correlator corresponding to the photon polarization $\vec\epsilon_{(1)}$, we proceed now with the correlators corresponding to $\vec\epsilon_{(2)}$. 

The procedure to find them is clearer when carried out in terms of the gauge invariant fields $E_i\equiv\partial_iA_0-\partial_0A_i$. Equations (\ref{eom1})-(\ref{eom3}) can be rewritten in terms of $E_i$ with the aid of the constraint
\be
-\frac{1}{{\cal B}{\cal F}}A_t'+\sin\vartheta\, A_x' + \cos\vartheta\,  e^{\phi}A_z'=0,\label{ueom}
\ee
resulting in the set of equations
\begin{eqnarray}
&& 
E_x''+\left[\left(\log \frac{e^{-\frac{3}{4}\phi}\sqrt{{\cal B}}{\cal F}}{u}\right)'+\frac{k_x^2}{\bar{k}^2}(\log {\cal B}{\cal F})'\right]E_x'+\frac{\bar k^2}{{\cal F}}E_x+\frac{e^\phi k_xk_z}{\bar k^2}(\log {\cal B}{\cal F})'E_z'=0, \label{eomEx}\cr && \\
&& 
E_z''+\left[\left(\log \frac{e^{\frac{1}{4}\phi}\sqrt{{\cal B}}{\cal F}}{u}\right)'+\frac{e^\phi k_z^2}{\bar k^2}(\log e^{\phi}{\cal B}{\cal F})'\right]E_z'+\frac{\bar k^2}{{\cal F}}E_z+\frac{k_xk_z}{\bar k^2}(\log e^{\phi}{\cal B}{\cal F})'E_x'=0\,,\cr && 
\label{eomEz}
\end{eqnarray}
where $\bar k^2\equiv \frac{k_0^2}{{\cal F}{\cal B}}-k_x^2-e^{\phi}k_z^2$. Note that for $\vartheta=0$ (momentum along the $z$-direction) and $\vartheta=\pi/2$ (momentum along the $x$-direction) the equations for $E_x$ and $E_z$ decouple.

The action (\ref{boundactexp}) at $u=\epsilon$ can be written in terms of these fields as 
\begin{eqnarray}
S_\epsilon&=&-2 K_\mt{D7}\int
dt\, d\vec x\, \left\{\frac{e^{- \frac{3}{4} \phi}{\sqrt{\cal B}}{\cal F}}{u\, k_0^2\, \bar k^2}\left[ \left(\frac{k_0^2}{{\cal B}{\cal F}}-e^\phi k_z^2\right)E_xE_x'+\bar k^2E_yE_y' 
\right.\right.
\cr && \hskip 4cm
\left.\left.
 +e^\phi k_xk_z(E_xE_z)'+\left(\frac{k_0^2}{{\cal B}{\cal F}}- k_x^2\right)e^\phi E_zE_z'  \right]\right\}_{u=\epsilon}\,. \cr && 
 \label{boundE}
\end{eqnarray}
To perform the variations of the action, note that differentiating with respect to $A_{i=x,y,z}$ just gives $k_0$ times the result of differentiating with respect to $E_i$, and to differentiate with respect to $A_0$ we have to sum $\sum_i k_i \frac{\delta}{\delta E_i}$.

Since we need to take the limit $\epsilon\rightarrow 0$, let us compute the on-shell value of the integrand in (\ref{boundE}) using the near-boundary expansion of the metric. This expansion can be found in \cite{Mateos:2011tv} and is given by
\begin{eqnarray}
{\cal F}&=&1+\frac{11}{24}a^2 u^2+{\cal F}_4 u^4+\frac{7}{12} a^4 u^4 \log u+O(u^6),\label{Fboud}\\
{\cal B}&=&1-\frac{11 }{24}a^2 u^2+{\cal B}_4 u^4-\frac{7}{12} a ^4 u^4 \log u+O(u^6),\label{Bboud}\\
\phi&=&-\frac{a ^2 }{4}u^2+\left(\frac{1152 {\cal B}_4+121 a ^4 }{4032}\right)u^4-\frac{a ^4}{6}  u^4 \log u+O(u^6),\label{phiboud}
\end{eqnarray}
where ${\cal F}_4$ and ${\cal B}_4$ are coefficients undetermined by the asymptotic equations of motion that can nonetheless be extracted from the numerical solution. They turn out to be irrelevant for our purposes.

Using (\ref{Fboud})-(\ref{phiboud}) we can solve equations (\ref{eomEx}) and (\ref{eomEz})  perturbatively to find
\begin{eqnarray}
E_x&=&E_x^{(0)}+E_x^{(2)} \cos\vartheta\,  u^2 -\frac{1}{24} \left(\frac{3}{4} E_x^{(0)} k_0^2 \cos\vartheta +5 E_x^{(2)} \right)\cos\vartheta\, a^2  u^4 + O\left(u^6\right)\, ,\label{Exbond}\\
E_z&=&E_z^{(0)}-E_x^{(2)} \sin\vartheta\,  u^2 + E_z^{(4)}u^4 -\frac{\cos^2\vartheta\,  a ^2 k_0^2}{16} \left( E_x^{(0)}  \tan\vartheta + E_z^{(0)} \right) u^4\log u+O\left(u^6\right)\,,\label{Ezbound}\cr &&
\end{eqnarray}
where $E_x^{(0)},\,  E_z^{(0)},\,  E_x^{(2)}$, and $E_z^{(4)}$ are undetermined coefficients of the expansion, which can be extracted from the numerical solution. Using the above expressions, we can write the value of (\ref{boundE}) close to the boundary as
\be
S_\epsilon=-2K_\mt{D7}\int dt\, d\vec x\left[{\cal L}_1 + {\cal L}_2 + {\cal L}_3+\ldots+O\left(u^2\right)\right]_{u=\epsilon}
\,, \label{ExpboundE}
\ee
where
\begin{eqnarray}
{\cal L}_1&=&-\frac{3}{4} \sin^2\vartheta\,  {E_x^{(0)}}^2 -\frac{1}{4} \cos^2\vartheta\,  {E_z^{(0)}}^2 - \cos\vartheta\, \sin\vartheta\,  E_x^{(0)} E_z^{(0)}\, , \cr
{\cal L}_2&=&\frac{1}{3 k_0^2}\left[\frac{1+5\cos 2\vartheta}{\cos\vartheta} E_x^{(0)} E_x^{(2)} + \frac{48}{a^2}\tan\vartheta\,  E_x^{(0)} E_z^{(4)} -10 \sin \vartheta\,  E_z^{(0)} E_x^{(2)} +\frac{48}{a^2}E_z^{(0)} E_z^{(4)} \right]\,,
 \cr
{\cal L}_3&=&-\left(E_x^{(0)}\sin\vartheta + E_z^{(0)}\cos\vartheta\right)^2 \log u\,, \label{L3}
\end{eqnarray}
and the ellipsis stands for the terms in the $y$-components that have already been dealt with in the previous section.

We notice that ${\cal L}_3$ diverges as we take the $u\rightarrow 0$ limit, which implies that the boundary action is sensitive to the anomaly of the background found in \cite{Mateos:2011tv}. In principle we would need to holographically renormalize this action, but the key observation here is that according to (\ref{combination}) the contribution of this divergent term to the production of photons with polarization $\vec\epsilon_{(2)}$ is proportional to
\be
\cos^2\vartheta\frac{\delta^2{\cal L}_3}{{\delta E_x^{(0)}}^2}+\sin^2\vartheta\frac{\delta^2{\cal L}_3}{{\delta E_z^{(0)}}^2}-2\cos\vartheta\sin\vartheta\frac{\delta^2{\cal L}_3}{\delta E_x^{(0)}\delta E_z^{(0)}}=0\,,
\ee
and therefore vanishes identically. This means that we only need to consider the contributions to the correlators coming from ${\cal L}_1$ and ${\cal L}_2$, which are finite and turn out to be given by
\begin{eqnarray}
\frac{G^\mt{R}_{xx}}{-4 K_\mt{D7}}&=&
-\frac{3}{4}k_0^2\sin^2\vartheta+\frac{1+5\cos 2\vartheta}{3\cos\vartheta}\frac{\delta E_x^{(2)}}{\delta E_x^{(0)}}+\frac{16\tan\vartheta}{a^2}\frac{\delta E_z^{(4)}}{\delta E_x^{(0)}}\, , 
\label{cxx} \\
\frac{G^\mt{R}_{zz}}{-4 K_\mt{D7}}&=&
-\frac{1}{4}k_0^2\cos^2\vartheta-\frac{10 \sin\vartheta}{3} \frac{\delta E_x^{(2)}}{\delta E_z^{(0)}} +\frac{16}{a^2}\frac{\delta E_z^{(4)}}{\delta E_z^{(0)}},
\label{czz}  \\
\frac{G^\mt{R}_{xz}}{-2 K_\mt{D7}}&=&
-k_0^2\cos\vartheta\, \sin\vartheta+\frac{1+5\cos 2\vartheta}{3\cos\vartheta}\frac{\delta E_x^{(2)}}{\delta E_z^{(0)}} \cr 
&& \hskip 2cm+\frac{16\tan\vartheta}{a^2}\frac{\delta E_z^{(4)}}{\delta E_z^{(0)}}-\frac{10 \sin\vartheta}{3} \frac{\delta E_x^{(2)}}{\delta E_x^{(0)}} 
+\frac{16}{a^2}\frac{\delta E_z^{(4)}}{\delta E_x^{(0)}}\,.
\label{cxz}
\end{eqnarray}
Summing up these expressions according to (\ref{combination}), one finds the simple result
\be
\chi_{(2)}\equiv \epsilon^\mu_{(2)}\,  \epsilon^\nu_{(2)}\, \chi_{\mu\nu} =
16 K_\mt{D7}\, \mbox{Im}\left(\cos\vartheta\frac{\delta E_x^{(2)}}{\delta E_x^{(0)}}-\sin\vartheta\frac{\delta E_x^{(2)}}{\delta E_z^{(0)}}\right)\,.
\label{combofinal}
\ee

To determine how the coefficients $E_x^{(2)}$ and $E_z^{(4)}$ vary with respect of $E_x^{(0)}$ and $E_z^{(0)}$, it is convenient to write the solutions $E_{x}$ and $E_{z}$ as a vector. When solving the equation of motion for 
\be
{\bf E}=\left( \begin{array}{c}
E_x \\ E_z \end{array} \right)
\ee
near the horizon by a power series expansion $(u-\uh)^\nu$,  we obtain that the exponent $\nu$ for both components of this vector is the same as that for the $A_y$ mode, namely $\nu=\pm i \wn/2$. After imposing the infalling wave condition, the rest of the power series is linearly determined by the value of ${\bf E}$ at the horizon. Integrating from the horizon using any choice of such a vector would pick a particular solution to the equations of motion. These are linear and therefore their general solution can be written as a linear combination of any two linearly independent solutions ${\bf E_1}=(E_{x,1},\, E_{z,1})^T$ and ${\bf E_2}=(E_{x,2},\, E_{z,2})^T$.

\begin{figure}
\begin{center}
\begin{tabular}{cc}
\setlength{\unitlength}{1cm}
\hspace{-0.9cm}
\includegraphics[width=7cm]{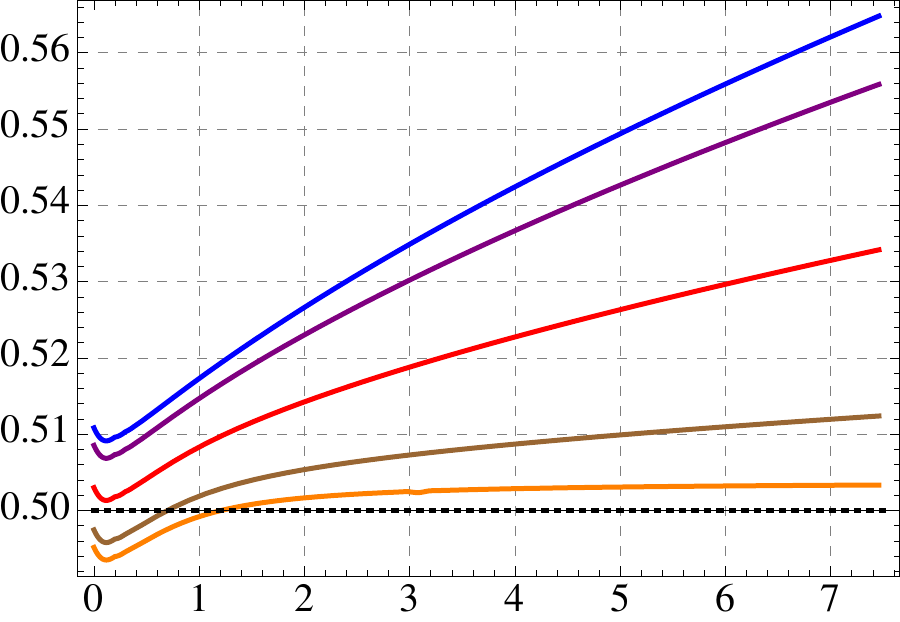} 
\qquad\qquad & 
\includegraphics[width=7cm]{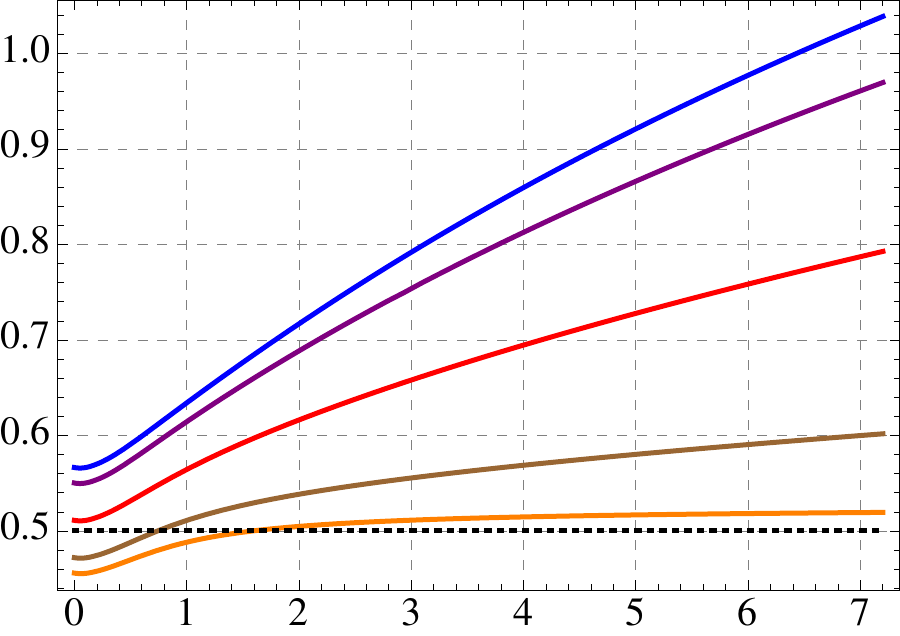}
\qquad
  \put(-455,40){\rotatebox{90}{$\chi_{(2)}/\chi_\mt{iso}(T)$}}
         \put(-250,-10){$\wn$}
         \put(-215,40){\rotatebox{90}{$\chi_{(2)}/\chi_\mt{iso}(T)$}}
         \put(-17,-10){$\wn$}
\\
(a) & (b)\\
& \\
\hspace{-0.9cm}
\includegraphics[width=7cm]{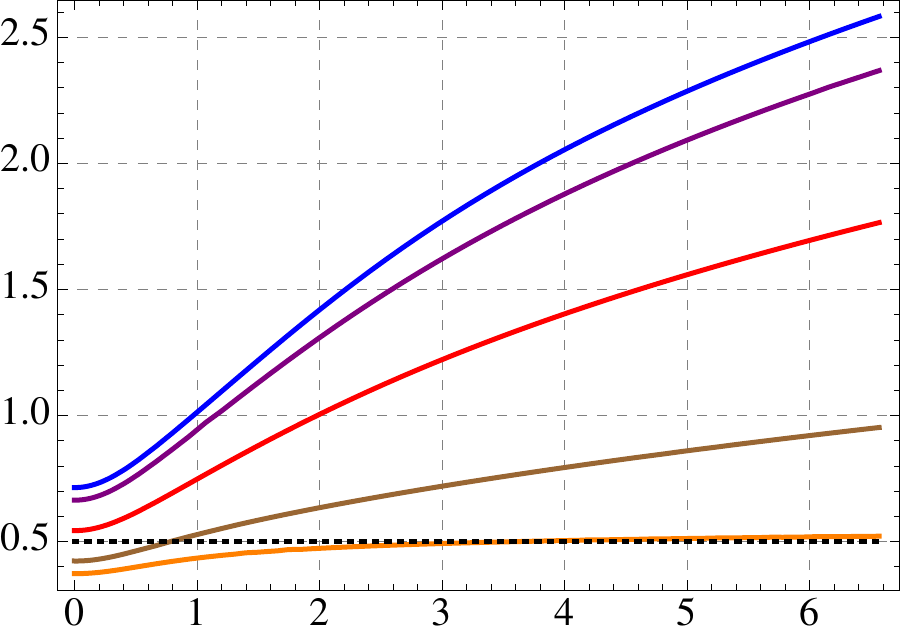} 
\qquad\qquad & 
\includegraphics[width=7cm]{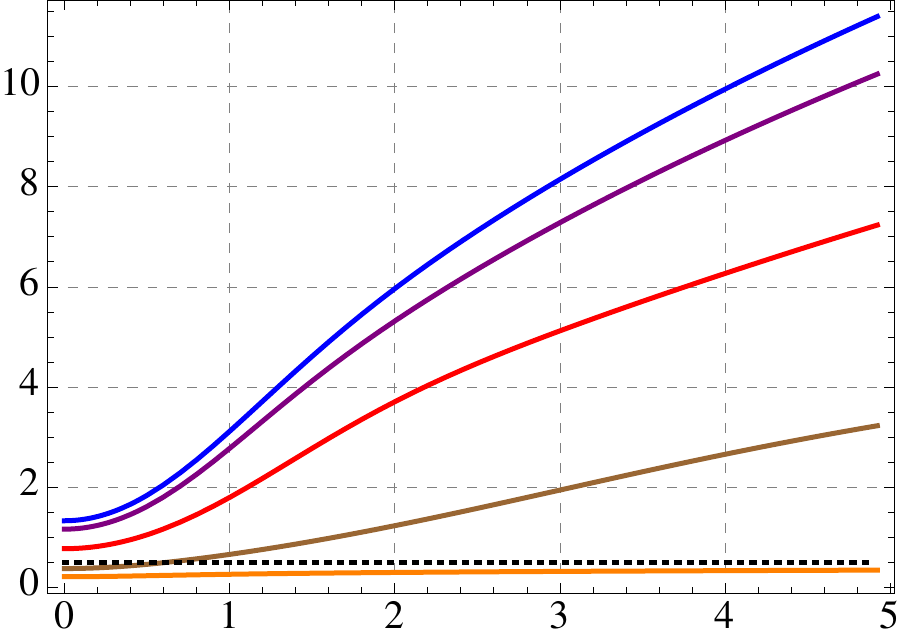}
\qquad
 \put(-455,40){\rotatebox{90}{$\chi_{(2)}/\chi_\mt{iso}(T)$}}
         \put(-250,-10){$\wn$}
         \put(-215,40){\rotatebox{90}{$\chi_{(2)}/\chi_\mt{iso}(T)$}}
         \put(-17,-10){$\wn$}
         \\
(c)& (d) 
\end{tabular}
\end{center}
\caption{\small Plots of the spectral density $\chi_{(2)}$ corresponding to the polarization $\epsilon_{(2)}$ normalized with respect to the isotropic result at fixed temperature $\chi_\mt{iso}(T)$. The curves correspond from top to bottom to the angles $\vartheta=0,\, \pi/8,\, \pi/4,\, 3\pi/8,\, \pi/2$. The four plots correspond to the cases $a/T=1.38$ (a), $4.41$ (b), $12.2$ (c), $86$ (d).}
\label{mixedT}
\end{figure}

\begin{figure}
\begin{center}
\begin{tabular}{cc}
\setlength{\unitlength}{1cm}
\hspace{-0.9cm}
\includegraphics[width=7cm]{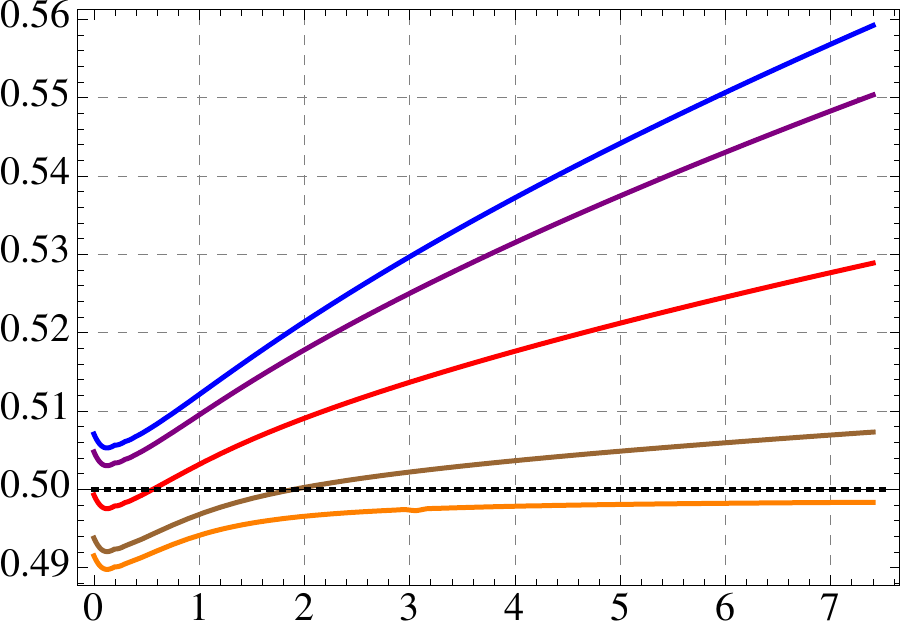} 
\qquad\qquad & 
\includegraphics[width=7cm]{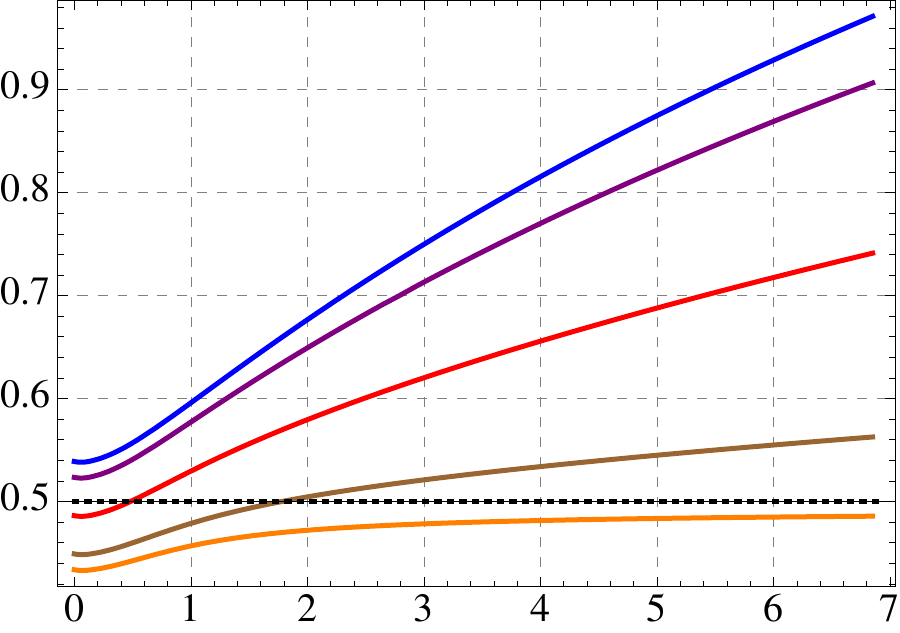}
\qquad
  \put(-455,40){\rotatebox{90}{$\chi_{(2)}/\chi_\mt{iso}(s)$}}
         \put(-250,-10){$\wn_s$}
         \put(-215,40){\rotatebox{90}{$\chi_{(2)}/\chi_\mt{iso}(s)$}}
         \put(-17,-10){$\wn_s$}
\\
(a) & (b)\\
& \\
\hspace{-0.9cm}
\includegraphics[width=7cm]{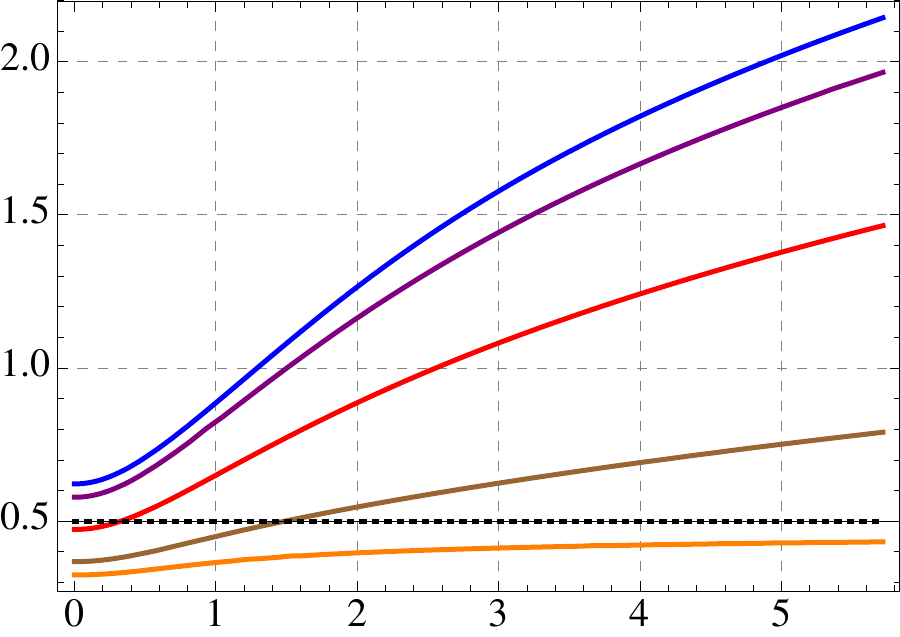} 
\qquad\qquad & 
\includegraphics[width=7cm]{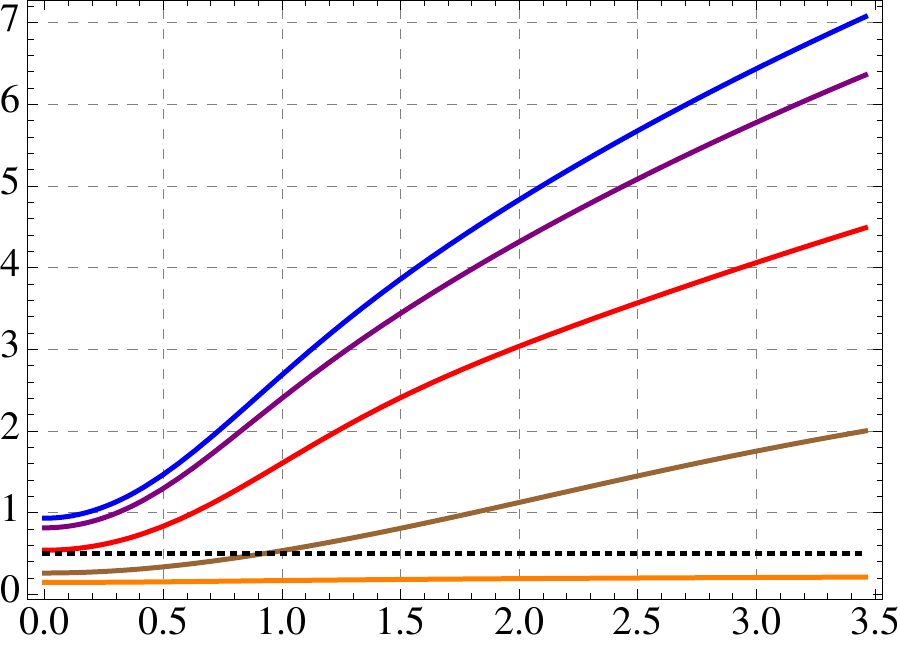}
\qquad
 \put(-455,40){\rotatebox{90}{$\chi_{(2)}/\chi_\mt{iso}(s)$}}
         \put(-250,-10){$\wn_s$}
         \put(-215,40){\rotatebox{90}{$\chi_{(2)}/\chi_\mt{iso}(s)$}}
         \put(-17,-10){$\wn_s$}
         \\
(c)& (d) 
\end{tabular}
\end{center}
\caption{\small Plots of the spectral density $\chi_{(2)}$ corresponding to the polarization $\epsilon_{(2)}$ normalized with respect to the isotropic result at fixed entropy density $\chi_\mt{iso}(s)$. The curves correspond from top to bottom to the angles $\vartheta=0,\, \pi/8,\, \pi/4,\, 3\pi/4,\, \pi/2$. The four plots correspond to the cases $a\nc^{2/3}/s^{1/3}=0.80$ (a), $2.47$ (b), $6.24$ (c), $35.5$ (d).}
\label{mixeds}
\end{figure}

Since we need to know how the solution depends on $E_x^{(0)}$ and $E_z^{(0)}$, the relevant base is the one formed by the two particular solutions that at the boundary reach the values
\be
{\bf E_1}\to \left(\begin{array}{c}1 \\ 0 \end{array}\right)\,, \qquad 
{\bf E_2}\to\left(\begin{array}{c}0 \\ 1 \end{array}\right), 
\ee
since in terms of this base the general solution can be written as 
\be
{\bf E}=E_x^{(0)} \, {\bf E_1}+E_z^{(0)} \, {\bf E_2}. \label{gensol}
\ee
Close to the boundary the general solution ${\bf E}$ will be described by the expansions (\ref{Exbond}) and (\ref{Ezbound}). From (\ref{gensol}) we see that its corresponding coefficients will be given by
\bea
 {E_x^{(2)}}=E_x^{(0)} \, {E_{x,1}^{(2)}} + E_z^{(0)}\,  {E_{x,2}^{(2)}}\,, \qquad 
 {E_z^{(4)}}=E_x^{(0)}\,  {E_{z,1}^{(4)}} + E_z^{(0)}\,  {E_{z,2}^{(4)}}, 
\label{coeffB}
\eea
where ${E_{x,i}^{(2)}}$ and ${E_{z,i}^{(4)}}$ (with $i=1,2$) are the coefficients for the expansions of ${\bf E_1}$ and ${\bf E_2}$ close to the boundary. From (\ref{coeffB}) it follows, for instance, that
\be
\frac{\delta {E_x^{(2)}}}{\delta E_x^{(0)}}={E_{x,1}^{(2)}} \,\qquad 
\frac{\delta {E_x^{(2)}}}{\delta E_z^{(0)}}={E_{x,2}^{(2)}}\,,
\label{varE0}
\ee
which can now be used in (\ref{combofinal}). In practice, since we can only solve the equations of motion numerically, we find any two linearly independent solutions, ${\bf E_{a}}=(E_{x,a},\, E_{z,a})^T$ and ${\bf E_{b}}=(E_{x,b},\, E_{z,b})^T$, to construct ${\bf E_1}$ and ${\bf E_2}$ as the linear combinations
\begin{eqnarray}
{\bf E_1}=
\frac{E^{(0)}_{z,b}\, {\bf E_{a}}-E^{(0)}_{z,a}\, {\bf E_{b}}}
{E^{(0)}_{x,a}\,  E^{(0)}_{z,b}- E^{(0)}_{x,b}\,  E^{(0)}_{z,a}} \,, 
\qquad 
{\bf E_2}=
\frac{E^{(0)}_{x,a}\, {\bf E_{b}}-E^{(0)}_{x,b}\, {\bf E_{a}}}
{E^{(0)}_{x,a} \, E^{(0)}_{z,b}- E^{(0)}_{x,b} \, E^{(0)}_{z,a}} \,,
\end{eqnarray}
whence we can see that
\begin{eqnarray}
E_{x,1}^{(2)}=
\frac{E^{(0)}_{z,b}\, E_{x,a}^{(2)}-E^{(0)}_{z,a}\,  E_{x,b}^{(2)}}
{E^{(0)}_{x,a}\,  E^{(0)}_{z,b}- E^{(0)}_{x,b}\,  E^{(0)}_{z,a}} \,, 
\qquad 
E_{x,2}^{(2)}=
\frac{E^{(0)}_{x,a}\,  E_{x,b}^{(2)}-E^{(0)}_{x,b}\,  E_{x,a}^{(2)}}
{E^{(0)}_{x,a} \, E^{(0)}_{z,b}- E^{(0)}_{x,b} \, E^{(0)}_{z,a}} \,.
\end{eqnarray}
With this ground work in place, we may now use these expressions to numerically obtain the coefficients we need to input in (\ref{combofinal}) to find $\chi_{(2)}$. The results are plotted in Figs.~\ref{mixedT} and \ref{mixeds}, where we compare, for different angles $\vartheta$, the photon production with polarization $\epsilon_{(2)}$ with the one of an isotropic plasma at the same temperature or at the same entropy density.
The $\vartheta=0$ curve for $\chi_{(2)}$ in all the plots matches the $\vartheta=0$ curve for $\chi_{(1)}$. This is expected, since in this case equations (\ref{eomy}) and (\ref{eom2}) are identical, being $g^{xx}=g^{yy}$. On the other hand, one does not expect a similar match between the $\vartheta=\pi/2$ curve for $\chi_{(2)}$ and the $\vartheta=\pi/2$ curve for $\chi_{(1)}$, since in this case equations (\ref{eomy}) and (\ref{eom3}) are different, with $g^{yy}\neq g^{zz}$, and indeed we obtain different curves. 

\begin{figure}
\begin{center}
\begin{tabular}{cc}
\setlength{\unitlength}{1cm}
\hspace{-0.9cm}
\includegraphics[width=7cm]{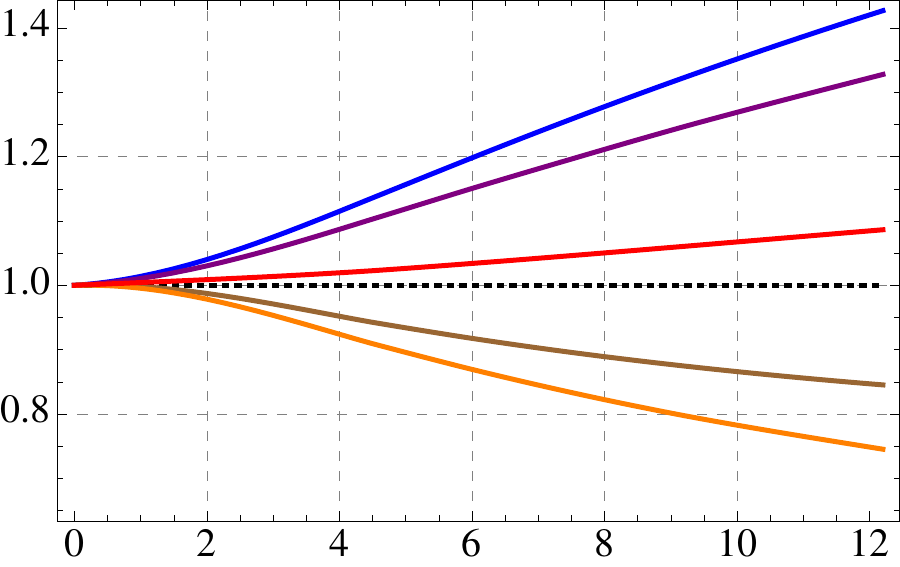} 
\qquad\qquad 
         & 
\includegraphics[width=7cm]{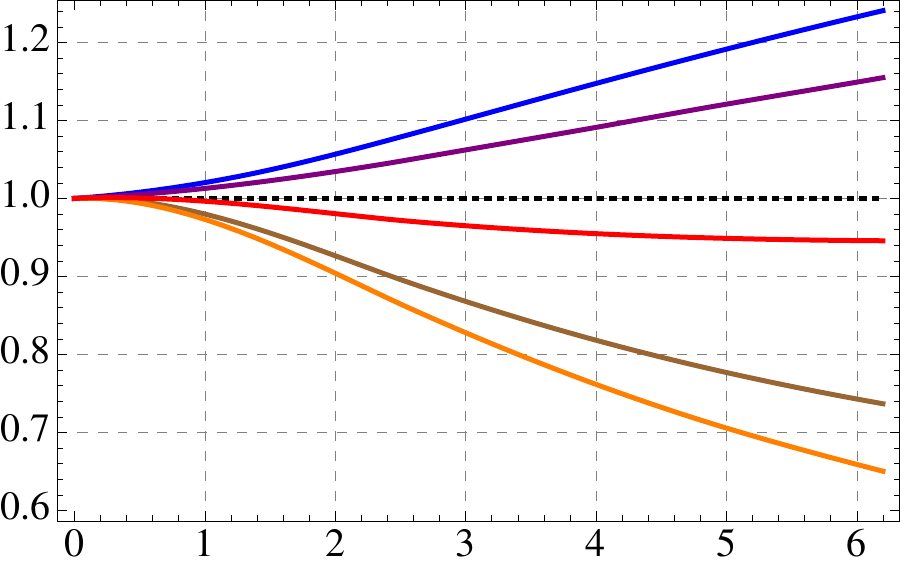}
\qquad
          \put(-455,80){\rotatebox{90}{$\sigma_{(2)}(T)$}}
         \put(-260,-15){$a/T$}
         \put(-217,80){\rotatebox{90}{$\sigma_{(2)}(s)$}}
         \put(-55,-15){$a\nc^{2/3}/s^{1/3}$}
         \\
         \end{tabular}
\end{center}
 \caption{\small Plot of the conductivity $\sigma_{(2)}$ corresponding to the polarization $\epsilon_{(2)}$ as a function of $a/T$ (left) and of $a\nc^{2/3}/s^{1/3}$ (right). The curves correspond from top to bottom to $\vartheta=0,\,\pi/8,\, \pi/4, \,3\pi/8,\,  \pi/2$.}
\label{mixed_conductivity_aoverT}
\end{figure}

\begin{figure}
    \begin{center}
        \includegraphics[width=0.6\textwidth]{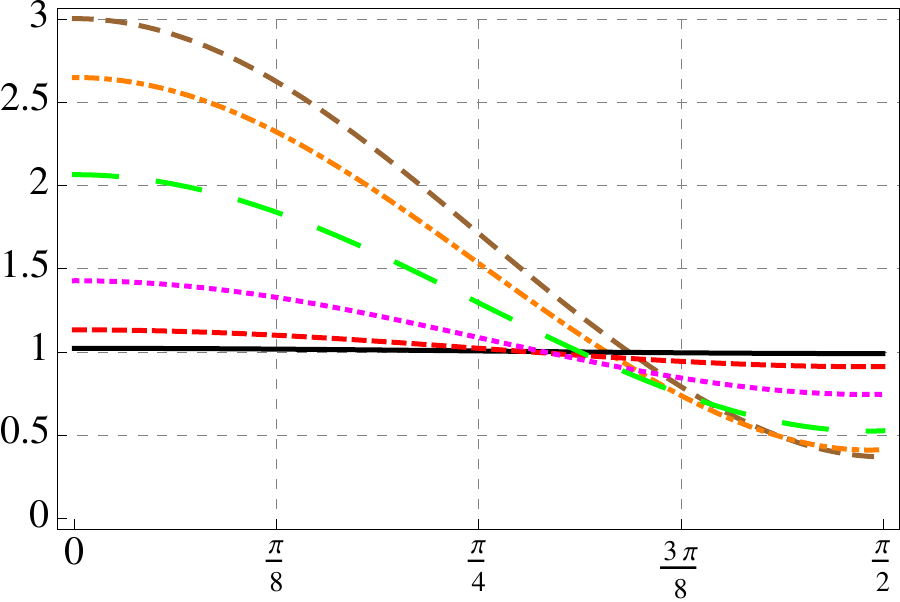}
         \put(-295,110){$\sigma_{(2)}(T)$}
         \put(10,10){$\vartheta$}
        \caption{\small Plot of the conductivity $\sigma_{(2)}$ corresponding to the polarization $\epsilon_{(2)}$ as a function of $\vartheta$. We plot the cases $a/T=1.38$ (black, solid curve),  $4.41$ (red, dashed curve), $12.2$ (magenta, dotted curve), $40$ (green, very coarsely dashed curve), $86$ (orange, dot-dashed curve), $120$ (brown, coarsely dashed curve).}
        \label{mixed-conductivity}
    \end{center}
\end{figure}

For photons with polarization along $\epsilon_{(2)}$, the conductivities
\bea
\sigma_{(2)}(T)&=&\lim_{k_0\to 0}\frac{\chi_{(2)}}{\chi_{(2),\mt{iso}}(T)}=\lim_{k_0\to 0}2\frac{\chi_{(2)}}{\chi_\mt{iso}(T)}\,,\cr
\sigma_{(2)}(s)&=&\lim_{k_0\to 0}\frac{\chi_{(2)}}{\chi_{(2),\mt{iso}}(s)}=\lim_{k_0\to 0}2\frac{\chi_{(2)}}{\chi_\mt{iso}(s)}
\eea
depend not only on the anisotropy $a$, as was the case for the polarization along the $y$-direction, but also on the angle $\vartheta$. Plots of $\sigma_{(2)}$ as a function of $a/T$ and $a\nc^{2/3}/s^{1/3}$ (for fixed $\vartheta$) and as a function of $\vartheta$ (for fixed $a/T$) are reported in Figs.~\ref{mixed_conductivity_aoverT} and \ref{mixed-conductivity}. For $\vartheta=0$ the conductivity is along the transverse $x$-direction (i.e. the direction of the polarization vector) and we recover the same results as (\ref{y_conductivity}) and $\sigma_\perp$ of \cite{rebhan_viscosity}. Similarly, for $\vartheta=\pi/2$ the conductivity is along the longitudinal $z$-direction and we recover $\sigma_z$ of \cite{rebhan_viscosity}.


\subsection{Total production rate}

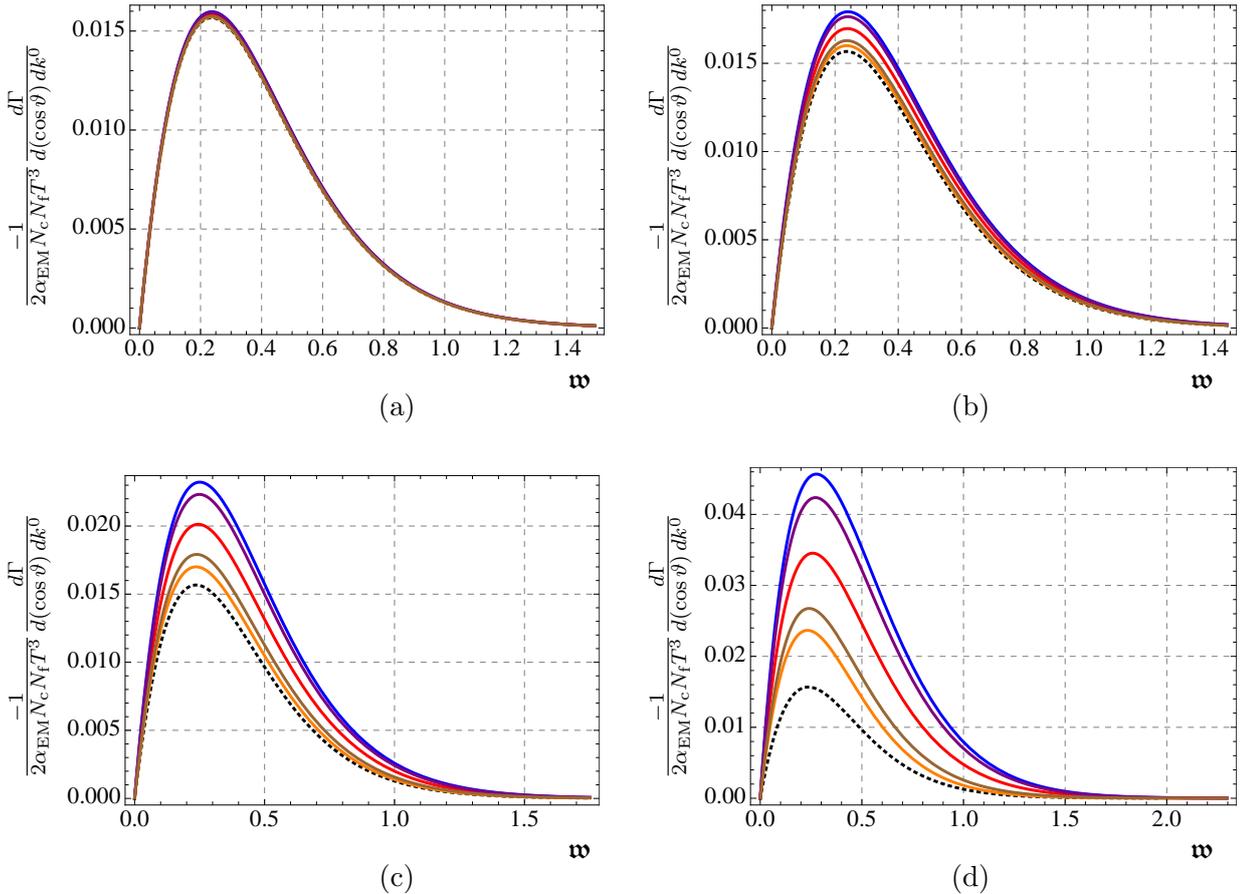
\begin{figure}
\begin{center}
\begin{tabular}{cc}
\setlength{\unitlength}{1cm}
\hspace{-0.9cm}
\includegraphics[width=7cm]{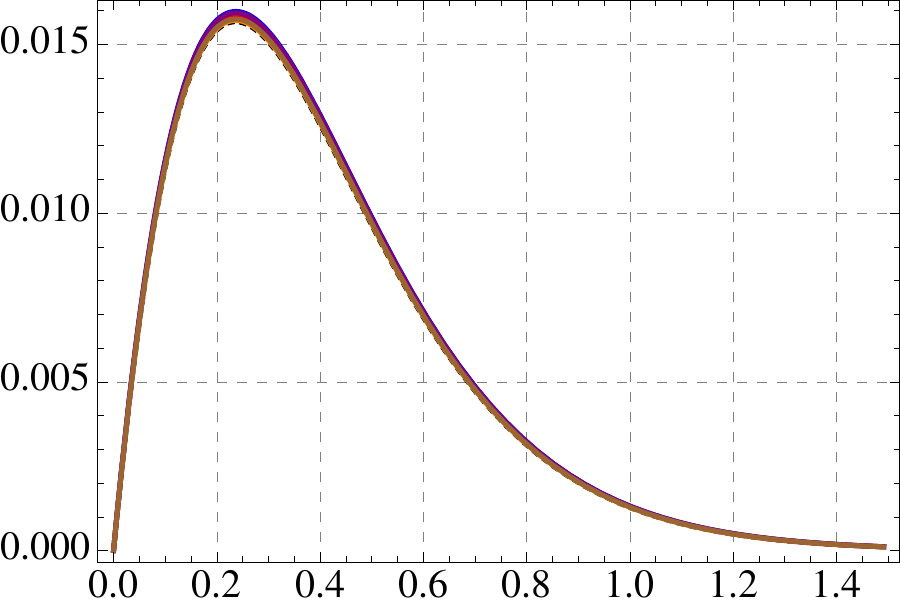} 
\qquad\qquad & 
\includegraphics[width=7cm]{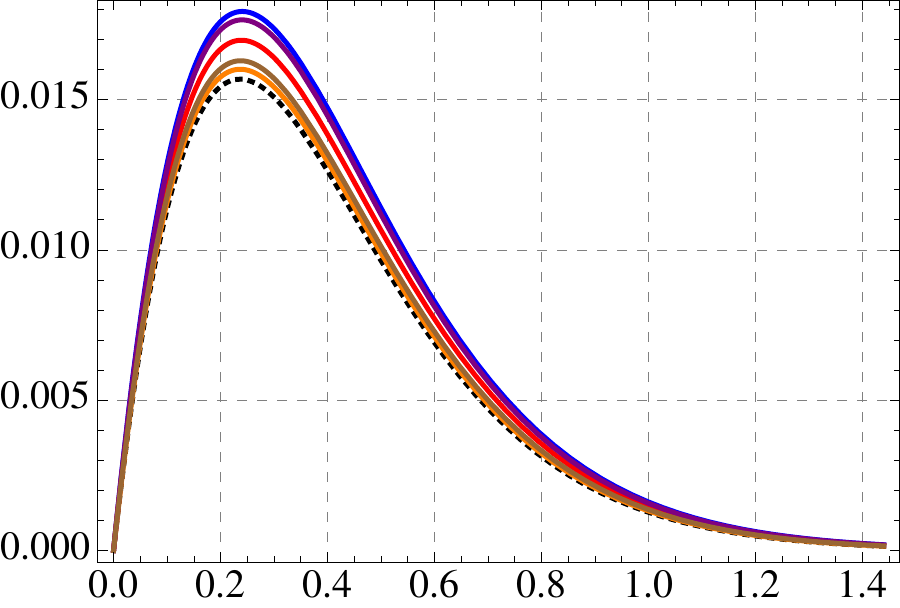}
\qquad
  \put(-460,20){\rotatebox{90}{$\frac{-1}{2\alpha_\mt{EM}\nc\nf T^3}\frac{d\Gamma}{d(\cos\vartheta)\, dk^0}$}}
         \put(-250,-10){$\wn$}
         \put(-220,20){\rotatebox{90}{$\frac{-1}{2\alpha_\mt{EM}\nc\nf T^3}\frac{d\Gamma}{d(\cos\vartheta)\, dk^0}$}}
         \put(-17,-10){$\wn$}
\\
(a) & (b)\\
& \\
\hspace{-0.9cm}
\includegraphics[width=7cm]{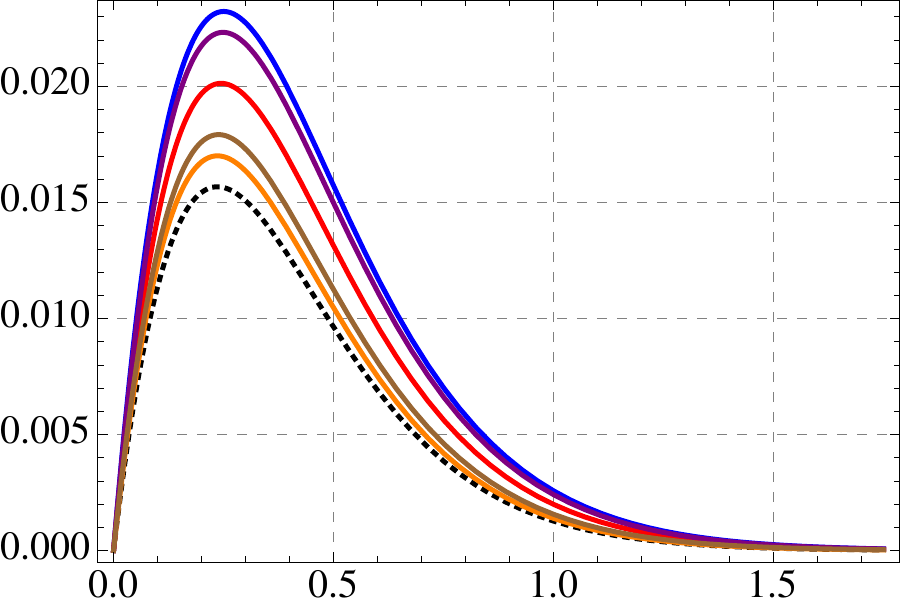} 
\qquad\qquad & 
\includegraphics[width=7cm]{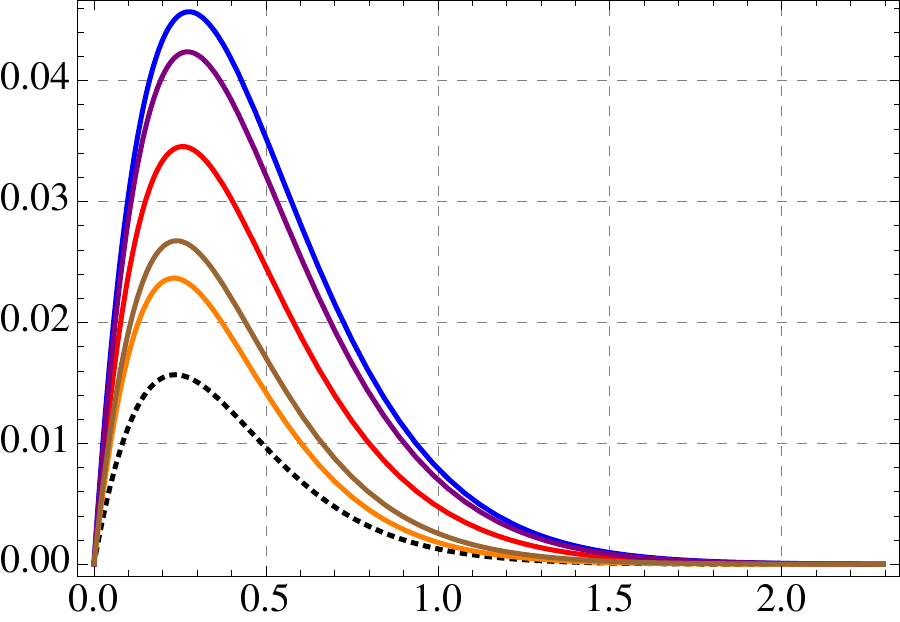}
\qquad
 \put(-460,20){\rotatebox{90}{$\frac{-1}{2\alpha_\mt{EM}\nc\nf T^3}\frac{d\Gamma}{d(\cos\vartheta)\, dk^0}$}}
         \put(-250,-10){$\wn$}
         \put(-220,20){\rotatebox{90}{$\frac{-1}{2\alpha_\mt{EM}\nc\nf T^3}\frac{d\Gamma}{d(\cos\vartheta)\, dk^0}$}}
         \put(-17,-10){$\wn$}
         \\
(c)& (d) 
\end{tabular}
\end{center}
\caption{\small Plots of the total production rate. The curves correspond from top to bottom to the angles $\vartheta=0,\, \pi/8,\, \pi/4,\, 3\pi/8,\, \pi/2$. The four plots correspond to the cases $a/T=1.38$ (a), $4.41$ (b), $12.2$ (c), $86$ (d). The temperatures in the four cases are, respectively, $T=0.32,\, 0.33,\, 0.36,\, 0.49$. The isotropic result at the same temperature is the dotted black curve. In (a) the curves are virtually indistinguishable from the isotropic one.}
\label{totalrate}
\end{figure}

We have now all the ingredients to calculate the total production rate (\ref{diff}). We convert this quantity to emission rate per unit photon energy in an infinitesimal angle around $\vartheta$. Using that the photon momentum is light-like, we have
\be
\frac{-1}{2\alpha_\mt{EM}\nc\nf T^3}\frac{d\Gamma}{d(\cos\vartheta)\, d k^0}=\frac{\wn}{2\nc\nf T^2}\frac{1}{e^{2\pi \wn}-1}\left(\chi_{(1)}+\chi_{(2)}\right)\,.
\ee
This quantity is plotted in Fig.~\ref{totalrate} for different values of $a/T$ and for different values of $\vartheta$. The isotropic result at the same temperature is \cite{Mateos:2007yp} 
\be
\frac{-1}{2\alpha_\mt{EM}\nc\nf T^3}\frac{d\Gamma_\mt{iso}}{d(\cos\vartheta)\, d k^0}=\frac{\wn^2}{2^\wn\left(e^{2\pi \wn}-1\right)
\left|{}_2F_1\left(1-\frac{1+i}{2}\wn_s,\, -\frac{1+i}{2}\wn_s,\, 1-i\wn_s;\, \frac{1}{2}\right)\right|^2}\,,
\ee 
and is shown as a black dotted curve in the figures.


\section{Discussion}
\label{sec4}

In this paper, we have computed the photon production rate and the electric conductivity of an ${\cal N}=4$ anisotropic plasma modeled holographically using the geometry of \cite{Mateos:2011ix,Mateos:2011tv}. The plasma is infinitely extended, in thermal equilibrium, strongly coupled and has a large number of colors $\nc$ and a quenched number of flavors, $\nf\ll\nc$. It is rotationally invariant in the $xy$-plane, but not in the $z$-direction, which we have called the anisotropic direction. In a heavy ion collision experiment, the latter would correspond to the beam direction, whereas the $xy$-plane would correspond to the transverse plane. We have considered arbitrary orientations of the photon momentum with respect to the anisotropic direction, as well as arbitrary values of the anisotropy parameter. 

The spectral densities are sensitive to the degree of anisotropy of the system and to the angle between the anisotropic direction and the photon wave vector, with the sensitivity increasing with the photon energy. This increasing sensitivity has also been found in anisotropic plasmas at weak coupling; see, e.g., \cite{weak1}. The $\wn\to 0$ limit of the spectral densities (i.e. the DC conductivity) is independent of the angle $\vartheta$ for photons with polarization $\epsilon_{(1)}$ (along the transverse $y$-direction) and dependent on $\vartheta$ for photons with polarization $\epsilon_{(2)}$ (in the $xz$-plane), as expected on general grounds.

A generic feature emerging from our study is that the spectral densities (and consequently the total production rate) are higher for photons with longitudinal wave vectors, $\vartheta=0$, than for the ones with transverse wave vectors, $\vartheta=\pi/2$.  The enhancement along the longitudinal direction is perhaps not surprising, since that is the preferred direction in the expansion of the plasma. This enhancement also takes place in the electric conductivity, which is larger (smaller) than the isotropic value along the perpendicular (transverse) direction, irrespectively of whether the comparison is made with an isotropic plasma at the same temperature or at the same entropy density. Note that the conductivity is in the direction of the electric field, which coincides with the direction of the polarization vector and not of the photon momentum. The total production rate depicted in Fig.~\ref{totalrate} is always larger than the isotropic value, for all frequencies, directions of propagation, and values of the anisotropy parameter. This means that the anisotropic plasma considered here always glows more brightly than its isotropic  counterpart.

The source of anisotropy in \cite{Mateos:2011ix,Mateos:2011tv} is a position-dependent theta-angle in the gauge theory, or, equivalently, a position-dependent axion in the gravity background. One might wonder how universal are the behaviors we have obtained, i.e. how they depend on the source of anisotropy used. We notice that the equations of motion for the gauge field on the branes' world-volume are solely determined by the metric and the dilaton. This implies that any other source of anisotropy that gives rise to a qualitatively similar metric and dilaton and no $B$-field will produce qualitatively similar results for the photon production rate,  irrespective of the other background fields (in particular, irrespective of whether or not there is a background axion turned on). 

Photon production in the strongly coupled anisotropic plasma of \cite{Janik:2008tc} has been worked out in \cite{Rebhan:2011ke}. The geometry in those papers presents a naked singularity, which nonetheless allows for infalling boundary conditions and the definition of a retarded Green function in the dual gauge theory. The small frequency behavior of the spectral densities found in \cite{Rebhan:2011ke} is totally different from the one found here, due perhaps to the absence of a {\it bona fide} horizon. In particular, they find that for any non-vanishing anisotropy, even if very small, the spectral densities go to zero faster than $\wn$, resulting in a vanishing DC conductivity. At large frequencies, on the other hand, their anisotropic results smoothly approach the isotropic limit when the anisotropic parameter is sent to zero, as in our case. They find that the photon production rate is enhanced (suppressed) with respect to the isotropic one in the forward direction and suppressed (enhanced) in the transverse direction if the anisotropy is prolate (oblate). This is to be contrasted with our results, where the rate is always enhanced with respect to the isotropic one, although it is generally larger in the forward direction than in the transverse one.

The geometry considered in the present paper is static and the corresponding plasma is in thermal equilibrium. This fact renders a comparison with a real-world plasma problematic, at least for small $\wn$, which corresponds to large times. The plasma evolves in fact from an initial out-of-equilibrium state toward a final isotropic state and the static approximation necessarily breaks down. The inclusion of time evolution in the present context would require a time-dependent geometry and a distribution function more general than the Bose-Einstein one. Finding such a solution might be somewhat simplified using the ideas of \cite{Heller:2012km}, but it is beyond the scope of the present paper. With this caveat in place, we can nonetheless try to compare our results with experimental data. To the best of our knowledge, the measurement of thermal photons has been performed only for rapidities up to 0.35, which correspond to angles up to approximately 20 degrees around the transverse direction. In this range, an enhancement of direct thermal photon production, as also observed in our Fig. \ref{totalrate}, has been reported for Au+Au collisions at RHIC \cite{experiment}. Detections at higher rapidities have only been used so far to determine the centrality of the collision and not to make photon production statistics.
 
Thinking about a possible comparison with experimental data, we also recall that the quarks considered here are massless. It would then be interesting to extend our analysis by including a non-vanishing mass for the quarks, thus bringing the model closer to QCD. A non-zero quark mass is achieved by considering embeddings of D7-branes that, unlike the case studied here, are not equatorial. The asymptotic distance between the D7-branes and the equatorial plane would determine the mass of the quarks in the dual gauge theory.\footnote{In a strongly coupled plasma there are no quasi-particles and the mass mentioned here would in fact simply be a microscopic parameter on which the physics depends, but without the direct interpretation of mass of a quasi-particle.} In the isotropic case, the photon production rate showed a dependence on this mass \cite{Mateos:2007yp}, so it is clear that such a dependence should also be expected in the anisotropic case. In particular, as the mass becomes large compared with the temperature or anisotropy, it should be possible to identify the appearance of highly localized resonances in the spectral functions, indicating the reconstitution of the mesons that were melted in the plasma \cite{Mateos:2007yp}. Understanding the role played by the anisotropy in this transition is certainly something worth exploring.  

Finally, another possible extension would be the computation of the production rate of dileptons, which are also known to be a good probe of the properties of the emitting medium.


\section*{Acknowledgements}
It is a pleasure to thank David Mateos for collaboration at the initial stages of this project and for helpful discussions. We also thank Alejandro Ayala and Gary Horowitz for comments. We are supported by DGAPA-UNAM grant IN117012 (LP) and by CNPq and DE-FG02-95ER40896 (DT).


\end{document}